\RequirePackage{fix-cm}
\documentclass[smallextended]{svjour3wide} % onecolumn (wide varsion)
\smartqed  % flush right qed marks, e.g. at end of proof
\usepackage{graphicx}
\usepackage{amsmath}
\usepackage{amssymb}
\usepackage{mathrsfs}
\usepackage{color}
\usepackage{bm}
\usepackage{ulem}
\DeclareMathOperator{\Tr}{Tr}
\newcommand{\BigFig}[1]{\parbox{12pt}{\Huge #1}}
\newcommand{\BigZero}{\BigFig{0}}
\journalname{Journal of Statistical Physics}
\begin{document}

\title{Surface Critical Phenomena of a Free Bose Gas with Enhanced Hopping at the Surface}
% \subtitle{Do you have a subtitle?\\ If so, write it here}
\titlerunning{Surface Critical Phenomena of a Free Bose Gas} % if too long for running head

\author{Hiroyoshi Nakano \and Shin-ichi Sasa}

\institute{
H. Nakano \and S. Sasa \at
Department of Physics, Kyoto University, Kyoto 606-8502, Japan \\
\email{h.takagi@scphys.kyoto-u.ac.jp, sasa@scphys.kyoto-u.ac.jp}
}

\date{Received: date / Accepted: date}

\maketitle

\begin{abstract}
 % 100 - 150 wards
We study the Bose--Einstein condensation in a tight-binding model with a hopping rate enhanced only on a surface. We show that this model exhibits two different critical phenomena depending on whether the hopping rate on the surface $t_s$ exceeds the critical value $5t/4$, where $t$ is the hopping rate in the bulk. For $t_s/t<5/4$, normal Bose--Einstein condensation occurs, while the Bose--Einstein condensation for $t_s/t\geq5/4$ is characterized by the spatial localization of the macroscopic number of particles at the surface. By exactly calculating the surface free energy, we show that for $t_s/t<5/4$, the singularity of the surface free energy stems from diverging the correlation length in the bulk, while for $t_s/t\geq5/4$, it is induced by the coupling effects between the bulk and surface.
% 4 - 6 keywords
\end{abstract}

\section{Introduction} \label{sec:intro}
%表面における特異的な振る舞い  濡れ転移ほか良く知られている現象
Surfaces are rarely the focus of thermodynamics research, and yet specific and interesting behaviors may be observed on a surface. Typical examples are wetting of the liquid on a solid wall~\cite{Cahn,Taborek,ECheng} and surface melting~\cite{J. F. van der Veen}. In particular, some phenomena related to surfaces can be understood in the context of critical phenomena. These phenomena have been understood employing theory and methods developed to investigate standard critical phenomena; e.g., mean field theory, scaling theory, the renormalization group method and Monte Carlo simulation~\cite{Binder1,Diehl1,Diehl2,Pleimling}.

%スピン系の先行研究:先行研究の少なさ
Two-dimensional surface systems exhibit more diverse critical phenomena than pure two-dimensional systems because bulk strongly affects the surface. In particular, near the bulk critical points, the bulk is nontrivially connected with the surface, and as the result, the surface displays complicated critical phenomena. These critical phenomena were discovered by the analysis of the semi-infinite Ising ferromagnets~\cite{Binder2,Binder3,Svrakic,Lubensky,Nakanishi,Moldover,Pohl}. However, beyond simple models such as $O(n)$ model and $\phi^4$ model~\cite{Wiener,Barber,Singh,Frohlich,Bray,Tsallis,Landau1,Peczak,Kikuchi,DiehlDietrich1,Deng,DiehlDietrich2}, it remains unclear how the surface connects to the bulk and what kind of critical phenomena are observed. The investigation of the surface critical phenomena remains to be studied beyond.

%表面とバルクの結びつきに注目する
In this work, we investigate the surface critical phenomena of an ideal Bose gas. Specifically, we study a tight-binding Bose gas with a hopping rate enhanced only on a surface. This model has two advantages. The first one is that this model exhibits nontrivial phase transitions, and the second one is that all quantities can be exactly calculated. By exactly calculating the single-particle energy eigenstates and the free energy, we examine how the surface and the bulk are connected in terms of the quantum mechanical nature and the thermodynamic nature. We stress that our analysis is performed by focusing on coupling effects between the bulk and the surface.

%BECの調査
Concretely, we demonstrate that this model exhibits two types of Bose--Einstein condensation depending on the hopping rate of the surface. When the hopping rate of the surface is below a certain critical value, this model exhibits the standard Bose--Einstein condensation. When the hopping rate of the surface exceeds a certain critical value, a new type of Bose--Einstein condensation occurs under the effect of the surface. This Bose--Einstein condensation involves the localization of the $O(L^3)$ particles to the first layer.  By analyzing Bose--Einstein condensation from the view of the condensation of $O(L^3)$ particles to the ground state, we reveal the influence of the surface on this phenomenon as follows. Let us increase the hopping rate of the surface from the value of the unenhanced case. When the hopping rate of the surface is below a critical value, there are no finite modification in the ground state in the large system size limit. When the hopping rate of the surface exceeds a critical value, the ground state finitely changes from the unenhanced case. The new ground state is two-dimensional bound state near the surface. This change of the ground state is understood as purely quantum mechanical nature. On the other hand, the enhanced hopping effect changes the thermodynamic nature of our system. When the temperature is sufficiently high, the change of the ground state does not lead to change of the physical quantities in the thermodynamic limit, since only $O(L^0)$ particles occupy the ground state. However, when the temperature is below a certain critical point and Bose--Einstein condensation occurs, $O(L^3)$ particles occupy the ground state. As the result, we directly observe the change of the ground state as the change of the physical quantities.

%自由エネルギーの調査
Furthermore, we examine two Bose-Einstein condensations in terms of the free energy. We notice that the total Helmholtz free energy $F(\beta,\rho)$ is expanded as
\begin{eqnarray}
F(\beta,\rho) = L^3 f(\beta,\rho) + L^2 f^s(\beta,\rho) + O(L),
\end{eqnarray}
where the system size is $L\times L\times L$, $\beta$ is the inverse temperature and $\rho$ is the particle number density. We refer to $f(\beta,\rho)$ as the bulk free energy density and $f^s(\beta,\rho)$ as the surface free energy unit per area. We clarify how the non-analyticity of $f^s(\beta,\rho)$ appears through connecting the bulk and the surface. For this purpose, we exactly calculate $f^s(\beta,\rho)$. Then, we show that in our model $f^s(\beta,\rho)$ is separated into two parts. The first part corresponds to the free energy density of the pure two-dimensional system with the chemical potential given by $\partial f(\beta,\rho)/\partial \rho$. The second part corresponds to the interaction between the degrees of the freedom in the surface through the bulk. The non-analyticity of $f^s(\beta,\rho)$ is induced by the competition between these two parts. For the standard Bose--Einstein condensation, the non-analyticity of $f^s(\beta,\rho)$ stems from the interaction through the bulk, which implies that the critical phenomena at the surface arise from diverging the correlation length in the bulk. When the hopping rate at the surface exceeds a certain critical value, the non-analyticity of $f^s(\beta,\rho)$ stems from coupling of the enhanced hopping effect at the surface with long-range correlation in the bulk. As the result, the type of the non-analyticity of $f^s(\beta,\rho)$ change at the critical value of the hopping rate at the surface.

The present work provides a simple picture that diverse critical phenomena at the surface may be caused by competition between pure two-dimensional effects and the interaction through the bulk. Although the phase transitions in this model is rather special, we expect that such structure of $f^s(\beta,\rho)$ is universal.

%先行研究との関連
As a related study, Robinson reported an ideal Bose gas with an attractive boundary condition from the standpoint of the bulk critical phenomenon~\cite{Robinson}. This model exhibits essentially the same Bose--Einstein condensation as our model. In this study, the boundary-condition dependence of physical properties was investigated~\cite{Landau2} and pathological behavior involving the order of the limit operation was reported~\cite{Lauwers}. The attractive boundary condition was introduced under a mathematical motivation and it is not clear how this boundary condition can be realized. By contrast, we introduce our model while considering a more realistic experimental system. In fact, Bose--Einstein condensation was experimentally realized in the optical lattice~\cite{Fallani,Peil,Jaksch}. The hopping rates between nearest neighbor sites relate to the lattice constant in the optical lattice, which may be controlled by choosing the wavelength of the optical lattice laser. Therefore, the parameters of our model are regarded as microscopic parameters that may be experimentally accessed in principle. Furthermore, we stress that the present work aims to exactly calculate the coupling effects between the bulk and the surface.

%この論文の構成
The remainder of this paper is organized as follows. Section~\ref{sec:Model} explains the setup of our model. We derive the quantum mechanical nature of our model in Section~\ref{sec:Quantum mechanical nature}. We find that our model exhibits Bose--Einstein condensation regardless of the enhanced strength on the surface in Section~\ref{sec:Bose-Einstein condensation}. We then demonstrate that the particle number at the surface becomes of macroscopic order when the hopping rate at the surface exceeds a critical value. The singularity type of the bulk free energy and that of the surface free energy are respectively studied in Sections~\ref{sec:Bulk free energy density} and \ref{sec:Surface free energy per unit area}. The final section is devoted to a brief summary and concluding remarks.

%%%%%%%%%%%%%%%%%%%%%%%%%%%%%%%%%
\section{Model} 
\label{sec:Model}
\subsection{Quantum mechanical setup and formulation}
We study a free Bose gas confined in a cubic box by considering a tight-binding model on a cubic lattice. We express the cubic lattice by
\begin{eqnarray}
\Lambda = \{(i_1,i_2,j) \in \mathbb{Z}^3 \ |\ 1 \leq i_1 \leq L\ ,\ 1 \leq i_2 \leq L ,\ 1 \leq j \leq L \}.
\label{eq:lattice}
\end{eqnarray}
The index $j$ denotes the vertical position and $\bm{i}=(i_1,i_2)$ represents the lattice position in the horizontal layer. We assume periodic boundary conditions in the horizontal directions and free surface boundary conditions in the vertical direction. The Hamiltonian of this system is written as
\begin{eqnarray}
\hat{H} &=& - \sum_{j=1}^{L} t_j^{\parallel} \sum_{\langle \bm{i},\bm{i}' \rangle} \Big( \hat{a}^{\dagger}_{\bm{i},j}  \hat{a}_{\bm{i}',j} + \hat{a}^{\dagger}_{\bm{i}',j}  \hat{a}_{\bm{i},j} \Big) - \sum_{j=1}^{L-1} t_j^{\perp} \sum_{\bm{i}} \Big( \hat{a}^{\dagger}_{\bm{i},j} \hat{a}_{\bm{i},j+1} + \hat{a}^{\dagger}_{\bm{i},j+1} \hat{a}_{\bm{i},j} \Big),
\label{eq:Hamiltonian_Original}
\end{eqnarray}
where $\langle \bm{i},\bm{i}' \rangle$ represents a nearest-neighbor pair in a horizontal layer. 
%The Hamiltonian is translational invariant in the horizontal directions.
$\hat{a}_{\bm{i},j}$ and $\hat{a}^{\dagger}_{\bm{i},j}$ are respectively the bosonic annihilation and creation operators, which obey the bosonic commutation relations
\begin{eqnarray}
& & [\hat{a}_{\bm{i},j},\hat{a}^{\dagger}_{\bm{i}',j'}] = \delta_{\bm{i}\bm{i}'} \delta_{jj'}, \\[3pt]
& & [\hat{a}^{\dagger}_{\bm{i},j},\hat{a}^{\dagger}_{\bm{i}',j'}] = [\hat{a}_{\bm{i},j},\hat{a}_{\bm{i}',j'}] = 0.
\end{eqnarray}
We concentrate on the case
\begin{eqnarray}
t_j^{\parallel}=
\begin{cases}
t_s \equiv t(1+\Delta) & {\rm for} \ j=1, \\[3pt]
t & {\rm otherwise},
\end{cases}
\end{eqnarray}
\begin{eqnarray}
t_j^{\perp} = t & {\rm for} \ j=1,2,\cdots,L-1,
\end{eqnarray}
where $\Delta \geq0$ represents the enhanced hopping strength at the surface layer.

Owing to translational invariance in layer $j$, the Fourier momentum representation in the horizontal direction allows us to diagonalize the Hamiltonian. The transformation from the coordinate representation is expressed by 
\begin{eqnarray}
\hat{a}^{\dagger}_{\bm{i},j} &=& \frac{1}{L} \sum_{\bm{k}} \hat{a}^{\dagger}_{\bm{k},j} e^{-i \bm{k} \cdot \bm{i}}, \\
\hat{a}^{}_{\bm{i},j} &=& \frac{1}{L} \sum_{\bm{k}} \hat{a}_{\bm{k},j} e^{i \bm{k} \cdot \bm{i}},
\end{eqnarray}
\begin{eqnarray}
\hat{a}^{\dagger}_{\bm{k},j} = \frac{1}{L} \sum_{\bm{i}} \hat{a}^{\dagger}_{\bm{i},j} e^{i \bm{k} \cdot \bm{i}}, \\
\hat{a}^{}_{\bm{k},j} = \frac{1}{L} \sum_{\bm{i}} \hat{a}^{}_{\bm{i},j} e^{- i \bm{k} \cdot \bm{i}}.
\end{eqnarray}
Here $\bm{k}$ is defined as
\begin{eqnarray}
k_d \equiv \frac{2\pi}{L} n_d,
\label{eq:Definition k}
\end{eqnarray}
with
\begin{eqnarray}
n_d = \begin{cases}
-\frac{L-2}{2},\cdots,-2,-1,0,1,2,\cdots,\frac{L-2}{2},\frac{L}{2} & {\rm for} \  L:{\rm even}, \\[3pt]
-\frac{L-1}{2},\cdots,-2,-1,0,1,2,\cdots,\frac{L-3}{2},\frac{L-1}{2} & {\rm for} \ L:{\rm odd}, \\
\end{cases}
\end{eqnarray}
where $d=1,2$.
The Hamiltonian can then be written in the form
\begin{eqnarray}
\hat{H} = \sum_{j=1}^{L} \sum_{j'=1}^{L} \sum_{\bm{k}} \hat{a}^{\dagger}_{\bm{k},j} A_{jj'}(\bm{k}) \hat{a}_{\bm{k},j'}.
\label{eq:def_H_ws}
\end{eqnarray}
The $L \times L$ matrix $A(\bm{k})$ is given by
\begin{eqnarray}
A(\bm{k}) = 
\begin{pmatrix}
(1+\Delta) \omega(\bm{k}) & \;\;\; - t &  \;\;\; 0 &  &  & \\[3pt]
- t & \;\;\; \omega(\bm{k}) & \;\;\;- t & & & \\[3pt]
0 & \;\;\; - t &  \;\;\; \omega(\bm{k}) &\multicolumn{3}{c}{\raisebox{2.5ex}[0pt]{\BigZero}}  \\[3pt]
 & & & \ddots & & \\[3pt]
 & & & & \omega(\bm{k}) & \;\;\; - t \\[3pt]
 \multicolumn{4}{c}{\raisebox{1.5ex}[0pt]{\BigZero}} & - t & \;\;\; \omega(\bm{k}) 
\end{pmatrix},
\end{eqnarray}
where $\omega(\bm{k})$ is defined as
\begin{eqnarray}
\omega(\bm{k}) = -2t \sum_{d=1}^{2} \cos k_d.
\label{eq:Definition Omega(k)}
\end{eqnarray}
Note that $\omega(\bm{k})$ is the single-particle energy eigenvalue of the system cutting out only one layer with hopping constant $t$. 

Let $n_{\bm{k},j}$ be the number of particles occupying the single-particle state $(\bm{k},j)$. A complete set of bases for the Fock space $\mathcal{F}$ is represented by
\begin{eqnarray}
|\bm{n}\rangle \equiv \prod_{\bm{k},j} \frac{(\hat{a}^{\dagger}_{\bm{k},j})^{n_{\bm{k}},j}}{\sqrt{n_{\bm{k},j}!}} |0 \rangle,
\end{eqnarray}
where $\bm{n} = (n_{\bm{k},j})_{\bm{k},j}$.

%熱力学的な設定
\subsection{thermodynamic setup and formulation}
We focus on thermodynamic properties of the system that consists of $N$ Bose particles. In this paper, all calculations are carried out in the grand canonical ensemble with inverse temperature $\beta$ and chemical potential $\mu$. The ensemble average of an operator $\hat{Q}$ that is a function of $(\hat{a}_{i,\bm{j}},\hat{a}^{\dagger}_{i,\bm{j}})$ (or $(\hat{a}_{i,\bm{k}},\hat{a}^{\dagger}_{i,\bm{k}})$) is expressed as
\begin{eqnarray}
<\hat{Q}>_{\beta,\mu} &\equiv& e^{\beta J(\beta,\mu)} \Tr (\hat{Q} e^{-\beta (\hat{H}-\mu \hat{N})}) \nonumber \\[3pt]
&=& e^{\beta J(\beta,\mu)} \sum_{\bm{n}} \langle \bm{n} |\hat{Q} e^{-\beta (\hat{H}-\mu \hat{N})}|\bm{n} \rangle ,
\end{eqnarray}
where $\hat{N}$ is the total-particle-number operator
\begin{eqnarray}
\hat{N} \equiv \sum_{i} \sum_{\bm{k}} \hat{a}^{\dagger}_{i,\bm{k}} \hat{a}_{i,\bm{k}},
\end{eqnarray}
and $J(\beta,\mu)$ is the grand canonical free energy
\begin{eqnarray}
J(\beta,\mu) \equiv -\frac{1}{\beta} \log \Tr (e^{-\beta (\hat{H}-\mu \hat{N})}).
\label{eq:grand canonical free energy}
\end{eqnarray}

\section{Quantum mechanical nature}
\label{sec:Quantum mechanical nature}
This section summarizes the quantum mechanical nature. Let $\epsilon_n(\bm{k})$ and $|\lambda_n(\bm{k}) \rangle$ be the single-particle energy eigenvalue and single-particle energy eigenstate of the Hamiltonian $\hat{H}$, respectively. These satisfy the single-particle eigenvalue equation
\begin{eqnarray}
\hat{H} |\lambda_n (\bm{k}) \rangle = \epsilon_n(\bm{k}) |\lambda_n(\bm{k}) \rangle,
\label{eq:Eigenvalue Equation H}
\end{eqnarray}
where we label the quantum states by $(n,\bm{k})$. $\bm{k}$ corresponds to the Fourier momentum. The index $n$ is used following the rule
\begin{eqnarray} 
\epsilon_1(\bm{k}) \leq \epsilon_2(\bm{k}) \leq \epsilon_3(\bm{k}) \leq \cdots \leq \epsilon_{L}(\bm{k}).
\label{eq:ord_EE}
\end{eqnarray}
Equation (\ref{eq:Eigenvalue Equation H}) is equivalent to
\begin{eqnarray}
A(\bm{k}) \bm{v}_n(\bm{k}) = \epsilon_n(\bm{k}) \bm{v}_n(\bm{k}),
\label{eq:Eigenvalue Equation vector}
\end{eqnarray}
where $\bm{v}_n(\bm{k}) = (v_n^1(\bm{k}),v_n^2(\bm{k}),\cdots,v_n^L(\bm{k}))$ is the $L$-component vector. Using $\bm{v}_n(\bm{k})$, we express $|\lambda_n(\bm{k}) \rangle$ as
\begin{eqnarray}
|\lambda_n(\bm{k}) \rangle = \sum_{j=1}^L v_n^j(\bm{k}) \hat{a}^{\dagger}_{\bm{k},j} |0 \rangle.
\end{eqnarray}
For convenience, we restrict ourselves to wave numbers satisfying $\omega(\bm{k})<0$ in this section. See Appendix~\ref{appsec:A} for the other wave numbers $\omega(\bm{k}) \geq 0$.

\subsection{Unenhanced case}
We study the case $\Delta=0$. Equation (\ref{eq:Eigenvalue Equation vector}) is rewritten as
\begin{eqnarray}
B(\bm{k}) \bm{v}^0_n(\bm{k}) = \epsilon_v^{0}(\bm{k}) \bm{v}^0_n(\bm{k})
\end{eqnarray}
with 
the $L \times L$ matrix $B(\bm{k})$ 
\begin{eqnarray}
B(\bm{k}) = 
\begin{pmatrix}
\omega(\bm{k}) & \;\;\; - t &  \;\;\; 0 &  &  & \\[3pt]
- t & \;\;\; \omega(\bm{k}) & \;\;\;- t & & & \\[3pt]
0 & \;\;\; - t &  \;\;\; \omega(\bm{k}) &\multicolumn{3}{c}{\raisebox{2.5ex}[0pt]{\BigZero}}  \\[3pt]
 & & & \ddots & & \\[3pt]
 & & & & \omega(\bm{k}) & \;\;\; - t \\[3pt]
 \multicolumn{4}{c}{\raisebox{1.5ex}[0pt]{\BigZero}} & - t & \;\;\; \omega(\bm{k}) 
\end{pmatrix}.
\end{eqnarray}
The superscript 0 represents quantities for the unenhanced case $\Delta=0$. The equation can be solved using the translational invariance. The result is
\begin{eqnarray}
\epsilon_n^{0}(\bm{k}) = \omega(\bm{k}) - 2t \cos (\frac{n \pi}{L+1}),
\label{eq:Eigenvalue B}
\end{eqnarray}
\begin{eqnarray}
v^0_{n,j}(\bm{k}) = C_B \times \sin (\frac{j n\pi}{L+1}),
\label{eq:Eigenvector B}
\end{eqnarray}
where $n=1,2,3,\cdots,L$. $C_B$ is the normalization constant.

\subsection{Relation between $\epsilon_n^0(\bm{k})$ and $\epsilon_n(\bm{k})$}
$\epsilon_n(\bm{k})$ is given by a solution $z$ for the characteristic equation
\begin{eqnarray}
\det \Big[A(\bm{k}) - z E_L \Big] = 0,
\label{eq:Characteristic Equation A}
\end{eqnarray}
where $E_L$ is the $L\times L$ unit matrix. To estimate $\epsilon_n(\bm{k})$, we define $f(z;\bm{k})$ as 
\begin{eqnarray}
f(z;\bm{k}) \equiv \det \Big[A(\bm{k}) - z E_L \Big].
\end{eqnarray}
Using the cofactor expansion, we express $f(z;\bm{k})$ as
\begin{eqnarray}
f(z;\bm{k}) &=& \Big(-z +(1+\Delta) \omega(\bm{k}) \Big) \det \Big[B_{L-1} (\bm{k}) - z E_{L-1} \Big] - t^2 \det \Big[ B_{L-2} (\bm{k}) -z E_{L-2} \Big] \nonumber 
\\[3pt]
&=& \det \Big[ B_{L} (\bm{k}) - z E_L \Big] + \Delta \omega(\bm{k})  \det \Big[ B_{L-1} (\bm{k}) - z E_{L-1} \Big],
\label{eq:Determinant A}
\end{eqnarray}
where the size of matrix $B(\bm{k})$ is explicitly written as the subscript. Here $f(z;\bm{k})$ is an $L$-th polynomial of $z$ because
\begin{eqnarray}
\det \Big[ B_{L} (\bm{k}) - z E_L \Big] = \prod_{n=1}^{L} \Big[-z + \omega(\bm{k}) - 2t \cos \Big(\frac{n\pi}{L+1}\Big) \Big],
\label{eq:Determinant B}
\end{eqnarray}
where we have used (\ref{eq:Eigenvalue B}). 

\begin{figure}
\centering
\includegraphics[width=8cm]{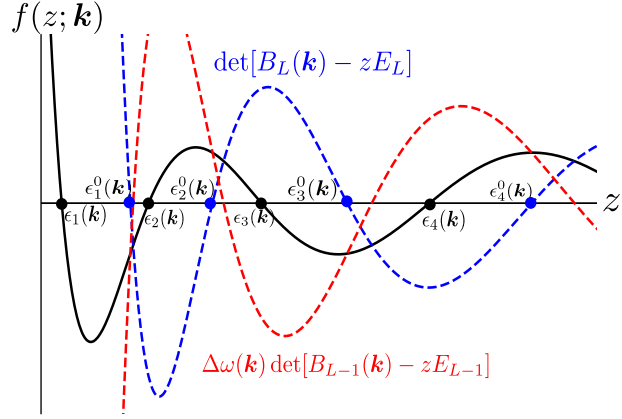}
\caption{Schematic graph of $f(z;\bm{k})$ (black), the first term of (\ref{eq:Determinant A}) (blue) and the second term of (\ref{eq:Determinant A}) (red) as a function of $z$ with $\bm{k}$ fixed. The zero points of $f(z;\bm{k})$ and the first term of (\ref{eq:Determinant A}) correspond to $\epsilon_n(\bm{k})$ and $\epsilon^0_n(\bm{k})$, respectively. }
\label{fig:f_zk}
\end{figure}

We associate $\epsilon_n(\bm{k})$ with $\epsilon_n^0(\bm{k})$ for given $(n,\bm{k})$. We notice that $f(z;\bm{k})$ is regarded as the sum of two oscillating functions. In Figure~\ref{fig:f_zk}, we present $f(z;\bm{k})$, the first term of (\ref{eq:Determinant A}) and the second term of (\ref{eq:Determinant A}) as a function of $z$ with $\bm{k}$ fixed. From Figure~\ref{fig:f_zk}, we find that $\epsilon_n(\bm{k})$ and $\epsilon^0_n(\bm{k})$ are one after the other. In more details, from (\ref{eq:Determinant A}) and (\ref{eq:Determinant B}), we obtain
\begin{eqnarray}
f(-\infty;\bm{k}) > 0
\label{eq:f(infty)}
\end{eqnarray}
and
\begin{eqnarray}
f(\epsilon_1^0(\bm{k});\bm{k}) &=& \Delta \omega(\bm{k}) \prod_{n=1}^{L-1} \biggl(2t \cos(\frac{\pi}{L+1})- 2t \cos(\frac{n\pi}{L}) \biggr) < 0.
\label{eq:f(eps1)}
\end{eqnarray}
From (\ref{eq:f(infty)}) and (\ref{eq:f(eps1)}), we obtain
\begin{eqnarray}
-\infty < \epsilon_1(\bm{k}) < \epsilon_1^0(\bm{k}).
\label{eq:rela_EE0}
\end{eqnarray}
Similarly, because
\begin{eqnarray}
f(\epsilon_2^0(\bm{k});\bm{k}) &=& \Delta \omega(\bm{k}) \prod_{n=1}^{L-1} \biggl(2t \cos(\frac{2\pi}{L+1})- 2t \cos(\frac{n\pi}{L}) \biggr) > 0,
\label{eq:f(eps2)}
\end{eqnarray}
we find
\begin{eqnarray}
\epsilon_1^0(\bm{k}) < \epsilon_2(\bm{k}) < \epsilon_2^0(\bm{k}).
\label{eq:rela_EE1}
\end{eqnarray}
Repeating similar procedures, we derive
\begin{eqnarray}
\epsilon_{n-1}^0(\bm{k}) < \epsilon_n(\bm{k}) < \epsilon_{n}^0(\bm{k})
\label{eq:rela_EE}
\end{eqnarray}
for $3 \leq n \leq L$.
Because these relations hold for any $L$, the energy eigenvalue $\epsilon_n(\bm{k})$ with $n \geq2$ cannot finitely deviate from $\epsilon_n^0(\bm{k})$ in the large system size limit, while there is no such restriction for the energy eigenvalue $\epsilon_1(\bm{k})$ as shown in (\ref{eq:rela_EE0}).

\subsection{$\epsilon_1(\bm{k})$ in the large system size limit}
\label{sec: stronger condition about eps1}
We derive a condition for $\epsilon_1(\bm{k})$ stronger than the relation (\ref{eq:rela_EE0}) in the large system size limit. While (\ref{eq:rela_EE0}) is true for all $L$, the bound in this section is valid only in the thermodynamic limit. We first rewrite (\ref{eq:Determinant B}) as
\begin{eqnarray}
\det \Big[B_{L} (\bm{k}) - z E_L \Big] &=& \exp \biggl\{L \sum_{n=1}^L  \log \Big( -z + \omega(\bm{k}) - 2t \cos (\frac{n\pi}{L+1}) \Big) \biggr\} \nonumber \\
&\simeq& \exp \biggl\{L \int_{0}^{1} dy \log \Big( -z + \omega(\bm{k}) - 2t \cos (\pi y) \Big) \biggr\}.
\end{eqnarray}
This integral converges when $-z+\omega(\bm{k})-2t>0$ and is calculated as
\begin{eqnarray}
\det \Big[B_{L} (\bm{k}) - z E_L \Big] \simeq  \frac{1}{2^L} \exp \biggl \{L \log \Big( \sqrt{(z-\omega(\bm{k}))^2-4t^2} -z+\omega(\bm{k}) \Big) \biggr \}.
\label{eq:Determinant B After Integral}
\end{eqnarray}
Recalling (\ref{eq:Determinant A}), we then express the characteristic equation (\ref{eq:Characteristic Equation A}) as 
\begin{eqnarray}
& & \exp \Bigg[(L-2)  \log \Big( \sqrt{(z-\omega(\bm{k}))^2-4t^2} -z+\omega(\bm{k}) \Big) \Bigg] \Bigg[\frac{1}{2} \Big(-z +(1+\Delta) \omega(\bm{k}) \Big) \nonumber
 \\[3pt]
&\times& \Big( \sqrt{(z -\omega(\bm{k}))^2- 4t^2} -z +  \omega(\bm{k}) \Big) - t^2 \Bigg]= 0
\label{eq:Characteristic Equation A After Integral}
\end{eqnarray}
for $z<\omega(\bm{k})-2t$. Because
\begin{eqnarray}
\exp \Bigg[(L-2)  \log \Big( \sqrt{(z-\omega(\bm{k}))^2-4t^2} -z+\omega(\bm{k}) \Big) \Bigg] > 0,
\end{eqnarray}
(\ref{eq:Characteristic Equation A After Integral}) is equivalent to 
\begin{eqnarray}
\frac{1}{2} \Big(-z +(1+\Delta) \omega(\bm{k}) \Big) \Big( \sqrt{(z -\omega(\bm{k}))^2- 4t^2} -z +  \omega(\bm{k}) \Big) - t^2 = 0.
\label{eq:Characteristic Equation A After Deformation}
\end{eqnarray}
For wave numbers satisfying
\begin{eqnarray}
\Delta \omega(\bm{k}) < - t,
\label{eq:k range solution exist}
\end{eqnarray}
we find the solution to (\ref{eq:Characteristic Equation A After Deformation}) with $z<\omega(\bm{k})-2t$ as
\begin{eqnarray}
z = \frac{t^2+\Delta^{2}\omega(\bm{k})^2}{\Delta \omega(\bm{k})} + \omega(\bm{k}).
\label{eq:solution k range solution exist}
\end{eqnarray}
Meanwhile, for wave numbers satisfying
\begin{eqnarray}
\Delta \omega(\bm{k}) \geq - t,
\label{eq:k range solution not exist}
\end{eqnarray}
there is no solution to the characteristic equation (\ref{eq:Characteristic Equation A}) for $z<\omega(\bm{k})-2t$. See Appendix~\ref{sec:solution of (42)} for the derivation.

Recalling that the energy eigenvalue satisfying $\epsilon_n(\bm{k}) < \omega(\bm{k})-2t$ corresponds to $\epsilon_1(\bm{k})$, we obtain
\begin{eqnarray}
\epsilon_1(\bm{k}) \to \frac{t^2+\Delta^{2}\omega(\bm{k})^2}{\Delta \omega(\bm{k})} + \omega(\bm{k})
\label{eq:eps1 in the large system size limit outside}
\end{eqnarray}
for $L \to \infty$, where $\bm{k}$ satisfies (\ref{eq:k range solution exist}). Meanwhile, for wave numbers satisfying (\ref{eq:k range solution not exist}), there is no energy eigenvalue in the range $\epsilon_n(\bm{k})<\omega(\bm{k})-2t$. Recalling (\ref{eq:rela_EE0}), we obtain a condition for $\epsilon_1(\bm{k})$ as
\begin{eqnarray}
\omega(\bm{k}) - 2t < \epsilon_1(\bm{k}) < \epsilon_1^0(\bm{k}).
\label{eq:stronger condition for eps1 in low Delta}
\end{eqnarray}
Note that $\epsilon_1(\bm{k})$ converges to $\epsilon_1^0(\bm{k})$ in the large system size limit.

\subsubsection{ground state}
Recalling (\ref{eq:Definition Omega(k)}), we find that the existence condition of the wave numbers satisfying (\ref{eq:k range solution exist}) is
\begin{eqnarray}
\Delta > \frac{1}{4}.
\end{eqnarray}
Note that the wave number $\bm{k}= \bm{0}$ satisfies condition (\ref{eq:k range solution exist}) for any $\Delta>1/4$. Therefore, with regard to the ground-state energy eigenvalue, (\ref{eq:eps1 in the large system size limit outside}) and (\ref{eq:stronger condition for eps1 in low Delta}) are written as
\begin{eqnarray}
\lim_{L \to \infty} \epsilon_1(\bm{0}) = 
\begin{cases}
- 6t & {\rm for} \ \Delta  \leq \frac{1}{4} ,\\[3pt]
 - 4 t - t \frac{1+16\Delta^2}{4\Delta} & {\rm for} \ \Delta  > \frac{1}{4}.\\
\end{cases}
\label{eq:ground state energy eigenvalue in the large L limit}
\end{eqnarray}

\subsection{Expression of $\bm{v}_n(\bm{k})$}
Let $\theta_n(\bm{k})$ be the $n$-th solution of the equation for $\theta$:
\begin{eqnarray}
-t \sin((L+1) \theta) - \Delta \omega(\bm{k}) \sin(L \theta) = 0.
\label{eq:eq_THETA}
\end{eqnarray}
Using $\theta_n(\bm{k})$, we derive $\epsilon_n(\bm{k})$ and $\bm{v}_n(\bm{k})$ as
\begin{eqnarray}
\epsilon_n(\bm{k}) = \omega(\bm{k}) -2t \cos \theta_n(\bm{k}),
\label{eq:eq_EE}
\end{eqnarray}
\begin{eqnarray}
v_n^j(\bm{k}) = \frac{v_n^1(\bm{k})}{\sin \theta_n(\bm{k})} \Big(\sin(j\theta_n(\bm{k})) + \frac{\Delta \omega(\bm{k})}{t} \sin((j-1)\theta_n(\bm{k})) \Big),
\label{eq:eq_U}
\end{eqnarray}
with $j=1,2,\cdots,L$ \cite{Cheng}. Here, we note that the order of the solutions of (\ref{eq:eq_THETA}) is given by that of the corresponding energy eigenvalues (\ref{eq:ord_EE}). In Appendix~\ref{sec:chech of vn}, we show that (\ref{eq:eq_EE}) and (\ref{eq:eq_U}) satisfy the eigenvalue equation (\ref{eq:Eigenvalue Equation vector}). Substituting (\ref{eq:eq_U}) into the normalization condition
\begin{eqnarray}
\sum_{j=1}^L |v_n^j(\bm{k})|^2 =1,
\label{eq:nor_U}
\end{eqnarray}
we obtain
\begin{eqnarray}
|v_n^1(\bm{k})|^2 = \frac{\sin^2 \theta_n(\bm{k})}{\sum_{j=1}^L \Big[\sin(j\theta_n(\bm{k})) +  \frac{\Delta \omega(\bm{k})}{t} \sin((j-1)\theta_n(\bm{k})) \Big]^2}.
\label{eq:eq_U1_2}
\end{eqnarray}

\subsubsection{$\Delta \leq \frac{1}{4}$}
From (\ref{eq:Definition Omega(k)}), (\ref{eq:rela_EE1}), (\ref{eq:rela_EE}) and (\ref{eq:stronger condition for eps1 in low Delta}), we find 
\begin{eqnarray}
- 6t < \epsilon_n(\bm{k}) < 6t
\label{eq:range of eps in low Delta}
\end{eqnarray}
for all $n$ and $\bm{k}$. From (\ref{eq:Definition Omega(k)}), (\ref{eq:eq_EE}) and (\ref{eq:range of eps in low Delta}), we obtain
\begin{eqnarray}
-1 < \cos \theta_n(\bm{k}) < 1.
\label{eq:range of cos theta in low Delta}
\end{eqnarray}
Because all $\theta_n(\bm{k})$ are real numbers, the denominator of (\ref{eq:eq_U1_2}) is evaluated as
\begin{eqnarray}
& & \sum_{j=1}^L\Big[\sin(j\theta_n(\bm{k})) + \frac{\Delta \omega(\bm{k})}{t} \sin((j-1)\theta_n(\bm{k})) \Big]^2 \nonumber \\[3pt]
&=& L \int_0^1 dx \Big[\sin(Lx\theta_n(\bm{k})) + \frac{\Delta \omega(\bm{k})}{t} \sin((Lx-1)\theta_n(\bm{k})) \Big]^2 + O(L^0) \nonumber \\[3pt]
&=& \frac{L}{2}\biggl(1 + \frac{\Delta^2 \omega(\bm{k})^2}{t^2} + 2 \frac{\Delta \omega(\bm{k})}{t}\cos \theta_n(\bm{k}) \biggr) + O(L^0) .
\end{eqnarray}
Substituting this result into (\ref{eq:eq_U1_2}), we obtain
\begin{eqnarray}
|v_n^1(\bm{k})|^2 \simeq \frac{2}{L} \frac{t^2 \sin^2 \theta_n(\bm{k})}{t^2 + \Delta^2 \omega(\bm{k})^2 + 2 t\Delta \omega(\bm{k})\cos \theta_n(\bm{k})}
\label{eq:res_U1_2}
\end{eqnarray}
in the large system size limit. Using (\ref{eq:eq_U}) and (\ref{eq:res_U1_2}), we express all $\bm{v}_n(\bm{k})$ in terms of $\theta_n(\bm{k})$. 

\subsubsection{$\Delta > \frac{1}{4}$}
From (\ref{eq:Definition Omega(k)}), (\ref{eq:rela_EE1}), (\ref{eq:rela_EE}) and (\ref{eq:eps1 in the large system size limit outside}), we find 
\begin{eqnarray}
\begin{cases}
\epsilon_n(\bm{k}) < - 6t \ \ {\rm for} \ n=1 \ {\rm and} \ \Delta \omega(\bm{k}) < - t ,\\
- 6t < \epsilon_n(\bm{k}) < 6t \ \ {\rm otherwise}.
\end{cases}
\label{eq:range of eps in high Delta}
\end{eqnarray}
From (\ref{eq:Definition Omega(k)}), (\ref{eq:eq_EE}) and (\ref{eq:range of eps in high Delta}), we find that $\theta_1(\bm{k})$ satisfies
\begin{eqnarray}
\cos \theta_1(\bm{k}) > 1,
\end{eqnarray}
where $\bm{k}$ satisfies $\Delta \omega(\bm{k}) < - t$. This means that $\theta_1(\bm{k})$ is a complex number. For the other wave numbers, we find that $\theta_1(\bm{k})$ is a real number. Note that the complex solution to (\ref{eq:eq_THETA}), $\theta_1(\bm{k})$, corresponds to the energy eigenvalue expressed by (\ref{eq:eps1 in the large system size limit outside}).

When $\theta_1(\bm{k})$ is a complex number, (\ref{eq:res_U1_2}) does not hold. Instead we can calculate $|v_n^1(\bm{k})|^2$ using the path integral expression. As shown in Appendix~\ref{sec:derivation of (u1n2)}, we obtain
\begin{eqnarray}
|v_n^1(\bm{k})|^2 = \lim_{z \to  \epsilon_n(\bm{k})} \biggl\{(-z + \epsilon_n(\bm{k}))\frac{\det (B_{L-1}(\bm{k}) - z E_{L-1})}{\det (A(\bm{k}) - z E_{L})}) \biggr\}.
\label{eq:path integral calculation u1n2}
\end{eqnarray}
Substituting (\ref{eq:Determinant A}), (\ref{eq:Determinant B After Integral}) and (\ref{eq:eps1 in the large system size limit outside}) into (\ref{eq:path integral calculation u1n2}), we obtain
\begin{align}
|v_1^1(\bm{k})|^2 %&=& \lim_{z \to  \epsilon_n(\bm{k})} \biggl\{(-z + \epsilon_n(\bm{k}))\frac{\det (B_{L-1}(\bm{k}) - z E_{L-1})}{\det \Big[ B_{L} (\bm{k}) - z E_L \Big] + \Delta \omega(\bm{k})  \det \Big[ B_{L-1} (\bm{k}) - z E_{L-1} \Big]}) \biggr\} \nonumber \\[3pt]
%&=& \lim_{z \to  \epsilon_n(\bm{k})} \biggl\{(-z + \epsilon_n(\bm{k}))\frac{1}{\frac{1}{2}\Big( \sqrt{(z-\omega(\bm{k}))^2-4J^2} -z+\omega(\bm{k}) \Big) + \Delta \omega(\bm{k})}) \biggr\} \nonumber \\[3pt]
= \frac{1}{4 \Delta \omega(\bm{k})} \biggl\{\sqrt{\frac{(t^2+4\Delta^{2}\omega(\bm{k})^2)^2}{(4\Delta \omega(\bm{k}))^2}-t^2} - \frac{t^2+4\Delta^{2} \omega(\bm{k})^2}{4\Delta \omega(\bm{k})} + 2\Delta \omega(\bm{k})\biggr\}
\label{eq:res_U1_2_c}
\end{align}
in the large size limit, where $\bm{k}$ satisfies (\ref{eq:k range solution exist}). Using (\ref{eq:eq_U}) and (\ref{eq:res_U1_2_c}), we express $\bm{v}_1(\bm{k})$ in terms of $\theta_1(\bm{k})$. When $\theta_1(\bm{k})$ is a real number, the previous result (\ref{eq:res_U1_2}) holds. From (\ref{eq:Definition Omega(k)}), (\ref{eq:eq_EE}) and (\ref{eq:range of eps in high Delta}), we find that $\theta_n(\bm{k})$ is a real number and the previous result (\ref{eq:res_U1_2}) holds for $n \geq 2$. We then express all $\bm{v}_n(\bm{k})$ in terms of $\theta_n(\bm{k})$.

$|v_n^1(\bm{k})|^2$ represents the existence probability of particle at the surface layer in $n$th eigenstate. From (\ref{eq:res_U1_2_c}), we find that $|v_1^1(\bm{k})|^2$ is $O(L^0)$ for $\Delta > 1/4$, which means that the particle is bound at the surface. In contrast, (\ref{eq:res_U1_2}) shows that $|v_1^1(\bm{k})|^2$ is $O(L^{-1})$ for $\Delta \leq 1/4$. Therefore, we find that this model exhibits the transition of the ground state from the homogeneous state to the bound state at $\Delta=1/4$. We note that this transition is induced by the enhanced hopping effects at the surface.

\section{Bose--Einstein condensation} 
\label{sec:Bose-Einstein condensation}
This section demonstrates that the model exhibits Bose--Einstein condensation. More precisely, above a critical density or below a critical temperature, the occupation number of the single-particle ground state is found to be of order $L^3$ regardless of $\Delta$.

Let $\hat{n}_n(\bm{k})$ be the operator of the occupation number of the single-particle state $(n,\bm{k})$ corresponding to the energy eigenvalue $\epsilon_n(\bm{k})$. The grand canonical average of this quantity is given as
\begin{eqnarray}
<\hat{n}_{n}(\bm{k})>_{\beta,\mu} = \frac{1}{e^{\beta (\epsilon_n(\bm{k})-\mu)}-1}.
\label{eq:Bose distribution function}
\end{eqnarray}
The chemical potential satisfies the condition
\begin{eqnarray}
\mu < \epsilon_1(\bm{0}),
\label{eq:range_mu}
\end{eqnarray}
so that the mean occupation number $<\hat{n}_{n}(\bm{k})>$ is not negative for any $\epsilon_n(\bm{k})$.

The total number of particles $N$ is obtained by the summation of the occupation number over all states:
\begin{eqnarray}
N = \sum_{n=1}^{\infty} \sum_{\bm{k}} \frac{1}{e^{\beta (\epsilon_n(\bm{k})-\mu)}-1}.
\end{eqnarray}
In particular, the particle number density $\rho$ is given as
\begin{eqnarray}
\rho = \frac{1}{L^3} \sum_{n=1}^{\infty} \sum_{\bm{k}} \frac{1}{e^{\beta (\epsilon_n(\bm{k})-\mu)}-1}.
\label{eq:eq_mu}
\end{eqnarray}
Solving this equation in $\mu$, we obtain the chemical potential $\mu = \mu(\beta,\rho)$ as a function of $(\beta,\rho)$. In the thermodynamic limit, (\ref{eq:eq_mu}) is expressed as
\begin{eqnarray}
\rho = \frac{1}{L^3} \frac{1}{e^{\beta(\epsilon_1(\bm{0})-\mu)}-1} + G(\beta,\mu)
\label{eq:eq_mu_gen}
\end{eqnarray}
with
\begin{eqnarray}
G(\beta,z) = \int \frac{d^3 \bm{\rho}}{(2\pi)^2 \pi} \frac{1}{e^{\beta(-2t \sum_{d=1}^3 \cos \rho_d - z)}-1} ,
\end{eqnarray}
where $z$ satisfies $-\infty < z \leq - 6t$. See Appendix~\ref{sec:derivation of (66)} for the derivation. The first term on the right-hand side of (\ref{eq:eq_mu_gen}) is the particle number density occupying the ground state and the second term represents the contribution of all excited states. 

\subsection{Derivation of Bose--Einstein condensation}
For (\ref{eq:eq_mu_gen}), we must consider that possible values of $\mu$ are restricted by condition (\ref{eq:range_mu}). Recalling (\ref{eq:ground state energy eigenvalue in the large L limit}), we find that this restriction can be classified into two cases depending on $\Delta$. In this subsection, the standard analysis for the Bose--Einstein condensation applies and only the saturation point of $\mu$ depends on the specific model.

\subsubsection{$\Delta < \frac{1}{4}$}
From (\ref{eq:rela_EE1}), (\ref{eq:rela_EE}) and (\ref{eq:stronger condition for eps1 in low Delta}), we find that no single-particle energy eigenvalue $\epsilon_n(\bm{k})$ finitely changes from the case $\Delta=0$ in the large system size limit. Because neither (\ref{eq:eq_mu_gen}) nor condition (\ref{eq:range_mu}) changes from that for $\Delta=0$, we observe the same critical behavior as for $\Delta=0$. To demonstrate this behavior, we consider the setting where $\rho$ is controlled by varying $\mu$, while $\beta$ is fixed.

\begin{figure}
\centering
\includegraphics[width=8cm]{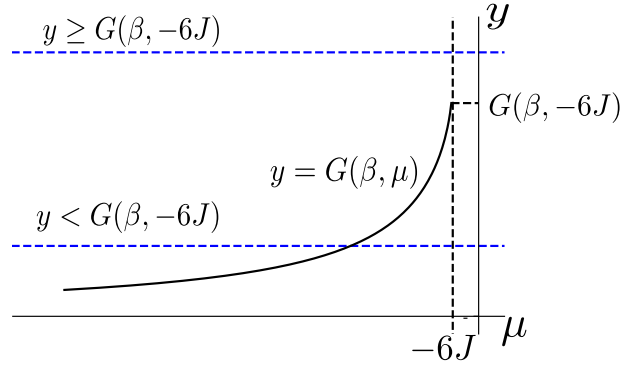}
\caption{Schematic graph of $G(\beta,\mu)$ as a function of $\mu$ with $\beta$ fixed. $G(\beta,\mu$) is defined in the region $\mu<-6t$. For any positive $\rho < G(\beta,-6t)$, there exists the unique $\mu$ satisfying $\rho=G(\beta,\mu)$, while there is no such $\mu$ that $\rho=G(\beta,\mu)$ for $\rho \geq G(\beta,-6t)$}.
\label{fig:G_betamu_andrhoP}
\end{figure}

We consider $G(\beta,\mu)$ as the function of $\mu$ with $\beta$ fixed. $G(\beta,\mu)$ is the monotonically increasing function of $\mu$ for any $\beta$. Then, $G(\beta,\mu) \leq G(\beta,-6t)$ because $\mu \leq -6t$. Figure \ref{fig:G_betamu_andrhoP} shows $G(\beta,\mu)$ as a function of $\mu$ at some fixed $\beta$.

According to the graph of $G(\beta,\mu)$, we consider the relation
\begin{eqnarray}
\rho = G(\beta,\mu).
\label{eq:eq_mu_HT}
\end{eqnarray}
For $\rho < G(\beta,-6t)$, we find a one-to-one correspondence between $\mu$ and $\rho$ through (\ref{eq:eq_mu_HT}). This pair $(\mu,\rho)$ also satisfies (\ref{eq:eq_mu_gen}) in the thermodynamic limit because
\begin{eqnarray}
\lim_{L \to \infty} \frac{1}{L^3} \frac{1}{e^{\beta(\epsilon_1(\bm{0})-\mu)}-1} = 0
\label{eq:limit 1}
\end{eqnarray}
holds. We therefore obtain $\mu$ satisfying (\ref{eq:eq_mu_gen}) as a function of $\rho$, which is expressed as
\begin{eqnarray}
\mu = G^{-1}(\beta,\rho)
\label{eq:eq_mu_HT after deform}
\end{eqnarray}
for $\rho < G(\beta,-6t)$, where $G^{-1}(\beta,\rho)$ is the inverse function of $G(\beta,\mu)$ as a function of $\mu$ with $\beta$ fixed.

Meanwhile, for $\rho \geq G(\beta,-6t)$, there is no $\mu$ satisfying (\ref{eq:eq_mu_HT}) (See Fig.~\ref{fig:G_betamu_andrhoP}). However, we find a one-to-one correspondence between $\mu$ and $\rho$ employing equation (\ref{eq:eq_mu_gen}) because the first term on the right-hand side of (\ref{eq:eq_mu_gen}) diverges when $\mu \to -6t$. We thus consider (\ref{eq:eq_mu_gen}) under the condition
\begin{eqnarray}
\beta(\epsilon_1(\bm{0})-\mu) \ll 1.
\label{eq:mu region}
\end{eqnarray}
We then estimate the right-hand side of (\ref{eq:eq_mu_gen}) in this region as
\begin{eqnarray}
\rho = \frac{1}{L^3} \frac{1}{\beta (\epsilon_1(\bm{0})-\mu)} + G(\beta,-6t) + o\Big(\frac{1}{L^3}\Big).
\end{eqnarray}
As a result, we obtain $\mu$ satisfying (\ref{eq:eq_mu_gen}) as a function of $\rho$:
\begin{eqnarray}
\mu = \epsilon_1(\bm{0}) - \frac{1}{L^3} \frac{1}{\beta (\rho-G(\beta^{\ast},-6t))} + o\Big(\frac{1}{L^3}\Big)
\label{eq:res_mu_SDLT0}
\end{eqnarray}
for $\rho \geq G(\beta,-6t)$.
Note that, in the thermodynamic limit, because
\begin{eqnarray}
\mu=-6t,
\label{eq:res_mu_SDLT}
\end{eqnarray}
$\mu$ does not determine $\rho$ uniquely in this $\rho$ region.

Recalling that the first term on the right-hand side of (\ref{eq:eq_mu_gen}) is the particle number density occupying the ground state, we obtain
\begin{eqnarray}
\frac{<\hat{n}_1(\bm{0})>_{\beta,\rho}}{L^3} = \rho - G(\beta,-6t)
\end{eqnarray}
for $\rho \geq G(\beta,-6t)$, where $<\hat{n}_1(\bm{0})>_{\beta,\rho}$ is not the canonical average but is defined as
\begin{eqnarray}
<\hat{n}_{1}(\bm{0})>_{\beta,\rho} \equiv <\hat{n}_{1}(\bm{0})>_{\beta,\mu(\beta,\rho)}.
\end{eqnarray}
When $<\hat{n}_{1}(\bm{0})>_{\beta,\rho}/L^3$ is finite, Bose--Einstein condensation is identified. In this sense, we define the critical density $\rho_c$ as
\begin{eqnarray}
\rho_c = G(\beta,-6t).
\end{eqnarray} 
One may also consider the Bose--Einstein condensation for the setting where $\beta$ is controlled with $\rho$ fixed. In this setting, we define the critical inverse temperature $\beta_c$ as
\begin{eqnarray}
\rho = G(\beta_c,-6t).
\label{eq:def_cri_temp_LD}
\end{eqnarray} 
We then observe Bose--Einstein condensation in the low-temperature regime $\beta\geq\beta_c$.

\subsubsection{$\Delta \geq \frac{1}{4}$}
As shown in (\ref{eq:ground state energy eigenvalue in the large L limit}), the ground-state energy finitely changes from that for $\Delta=0$ in the large system size limit. This change leads to a different type of critical phenomenon. We consider the setting that $\rho$ is controlled through $\mu$, while $\beta$ is fixed. From (\ref{eq:ground state energy eigenvalue in the large L limit}) and (\ref{eq:range_mu}), we find that possible values of $\mu$ are restricted to
\begin{eqnarray}
\mu < \mu_{\ast},
\label{eq:mu_range_HD}
\end{eqnarray}
where we have introduced
\begin{eqnarray}
\mu_{\ast} = - 4 t - t \frac{1+16\Delta^2}{4\Delta} .
\end{eqnarray}
In this range, we consider (\ref{eq:eq_mu_gen}).

\begin{figure}
\centering
\includegraphics[width=8cm]{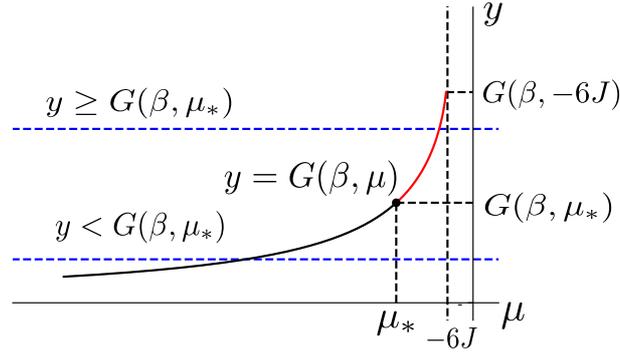}
\caption{Schematic graph of $G(\beta,\mu)$ as a function of $\mu$ with $\beta$ fixed in the region $\mu<\mu_{\ast}$. For any positive $\rho<G(\beta,\mu_{\ast})$, there exists a unique $\mu$ satisfying $\rho=G(\beta,\mu)$. Because we consider the restricted $\mu$ region, we find that there is no $\mu$ for $\rho\geq G(\beta,\mu_{\ast})$.}
\label{fig:G_betamu_andrhoP3}
\end{figure}

We first consider (\ref{eq:eq_mu_HT}) for $\rho < G(\beta,\mu_{\ast})$. This equation is identical to that for $\Delta<1/4$ because $\rho < G(\beta,\mu_{\ast})$ implies $\rho < G(\beta,- 6t)$ (See Fig. \ref{fig:G_betamu_andrhoP3}). Therefore, Repeating the same argument for $\Delta<1/4$, we obtain $\mu$ satisfying 
\begin{eqnarray}
\mu = G^{-1}(\beta,\rho)
\label{eq:eq_mu_HT after deform in HD}
\end{eqnarray}
for $\rho < G(\beta,\mu_{\ast})$.

We next consider (\ref{eq:eq_mu_HT}) for $\rho \geq G(\beta,\mu_{\ast})$. There is no $\mu$ satisfying (\ref{eq:eq_mu_HT}) in the range (\ref{eq:mu_range_HD}) (See Fig. \ref{fig:G_betamu_andrhoP3}). However, we find a one-to-one correspondence between $\mu$ and $\rho$ through (\ref{eq:eq_mu_gen}) because the first term of (\ref{eq:eq_mu_gen}) makes a finite contribution.
Considering (\ref{eq:eq_mu_gen}) under the condition
\begin{eqnarray}
\beta (\epsilon_1(\bm{0})-\mu) \ll 1,
\end{eqnarray}
we obtain $\mu$ satisfying (\ref{eq:eq_mu_gen}) as a function of $\rho$:
\begin{eqnarray}
\mu = \epsilon_1(\bm{0}) - \frac{1}{L^3} \frac{1}{\beta(\rho-G(\beta,\mu_{\ast}))} + o\Big(\frac{1}{L^3}\Big)
\end{eqnarray}
for $\rho \geq G(\beta,\mu_{\ast})$.
In the thermodynamic limit, we obtain
\begin{eqnarray}
\mu &=& \mu_{\ast}.
\label{eq:res_mu_HDLT}
\end{eqnarray}
The particle number density occupying the ground state is given as
\begin{eqnarray}
\frac{<\hat{n}_1(\bm{0})>_{\beta,\rho}}{L^3} = \rho - G(\beta,\mu_{\ast}) 
\label{eq:ground state occupation number_HD}
\end{eqnarray}
for $\rho \geq G(\beta,\mu_{\ast})$. The occupation density of the ground state is finite so that Bose--Einstein condensation occurs even for the case $\Delta>1/4$. We obtain the critical density $\rho_c$ as
\begin{eqnarray}
\rho_c = G(\beta,\mu_{\ast}) .
\label{eq:def_betac_HD}
\end{eqnarray}
The Bose--Einstein condensation for $\Delta \geq 1/4$ has unusual behavior such that the critical density $\rho_c$ depends on the value of $\Delta$ defined only at the boundary. Keeping this strong surface effect in mind, we refer to the behavior for $\Delta \geq 1/4$ as the Bose--Einstein condensation affected by the surface.

Similarly, when $\beta$ is controlled with $\rho$ fixed, we obtain the critical inverse temperature $\beta_c$ as
\begin{eqnarray}
\rho = G(\beta_c,\mu_{\ast}).
\label{eq:def_cri_temp_HD}
\end{eqnarray} 
In the low-temperature regime $\beta\geq\beta_c$, we observe the Bose--Einstein condensation state affected by the surface.

\subsection{Particle number near the surface}
The grand canonical average of the particle number in the $j$th layer $N_j$ is given by 
\begin{eqnarray}
N_j = <\sum_{\bm{k}} \hat{a}^{\dagger}_{j,\bm{k}} \hat{a}^{}_{j,\bm{k}}>_{\beta,\mu}.
\label{eq:def_Nj}
\end{eqnarray}
Using (\ref{eq:Bose distribution function}), we express $N_j$ in terms of the single-particle energy eigenvector $\bm{v}_n(\bm{k})$ as
\begin{eqnarray}
N_j &=& \sum_{\bm{k}} \sum_{n=1}^L |v_n^j(\bm{k})|^2  \frac{1}{e^{\beta(\epsilon_n(\bm{k})-\mu)}-1}.
\label{eq:Nj after calculation}
\end{eqnarray}
Below, we estimate $N_j$ for cases $\Delta<1/4$ and $\Delta \geq 1/4$ by calculating $|v_n^j(\bm{k})|^2$.

\subsubsection{$\Delta < \frac{1}{4}$}
From (\ref{eq:eq_U}) and (\ref{eq:res_U1_2}), we obtain
\begin{eqnarray}
|v^j_n(\bm{k})|^2 \simeq \frac{1}{L} \frac{\Big(t\sin (j \theta_n(\bm{k})) + \Delta \omega(\bm{k})\sin ((j-1)\theta_n(\bm{k}))\Big)^2}{t^2 + \Delta^2 \omega(\bm{k})^2 + 2 t\Delta \omega(\bm{k})\cos \theta_n(\bm{k})}.
\label{eq:unj}
\end{eqnarray}
Because $|v^j_n(\bm{k})|^2$ is $O(L^{-1})$, $N_j$ is $O(L^2)$. Such dependence on $L$ is the same as that for $\Delta=0$.

\subsubsection{$\Delta \geq \frac{1}{4}$}
We focus on the particle number in the first layer $N_1$. For $\bm{k}$ satisfying (\ref{eq:k range solution exist}), $|v^1_1(\bm{k})|^2$ is given by (\ref{eq:res_U1_2_c}), which is $O(1)$. For the other $\bm{k}$, $|v^1_1(\bm{k})|^2$ is given by (\ref{eq:unj}), which is $O(L^{-1})$. For the Bose--Einstein condensate, the contribution of the ground state to $N_1$ is
\begin{align}
L^3 \frac{1}{4 \Delta \omega(\bm{0})} \biggl\{\sqrt{\frac{(t^2+4\Delta^{2}\omega(\bm{0})^2)^2}{(4\Delta \omega(\bm{0}))^2}-t^2} - \frac{t^2+4\Delta^{2} \omega(\bm{0})^2}{4\Delta \omega(\bm{0})} + 2\Delta \omega(\bm{0})\biggr\}\times \Big( \rho - G(\beta,\mu_{\ast}) \Big),
\end{align}
which is $O(L^3)$. $N_1$ is therefore also $O(L^3)$. This $L$-dependence of $N_1$ is different from that for $\Delta<1/4$. This result implies that the Bose--Einstein condensate for $\Delta\geq 1/4$ is a spatially localized state in the first layer.

\section{Bulk free energy density}
\label{sec:Bulk free energy density}
This section investigates the free energy density for the two types of Bose--Einstein condensation focusing on the nonanalyticity of the free energy density. Here, the bulk free energy density of the grand canonical ensemble is defined by
\begin{eqnarray}
j(\beta,\mu;\Delta) &\equiv& \lim_{L\to\infty} \frac{1}{L^3} J(\beta,\mu;\Delta),
\label{eq:free energy density}
\end{eqnarray}
where $J(\beta,\mu;\Delta)$ is given by (\ref{eq:grand canonical free energy}). We first consider the $\Delta$-dependence of $j(\beta,\mu;\Delta)$. By straightforward calculation, the total free energy is expressed in terms of energy eigenvalues as
\begin{eqnarray} 
J(\beta,\mu;\Delta) = \frac{1}{\beta} \sum_{\bm{k}} \sum_{n=1}^{L} \log (1-e^{-\beta (\epsilon_n(\bm{k})-\mu)}).
\label{eq:J}
\end{eqnarray}
Taking the thermodynamic limit, we obtain
\begin{eqnarray}
\beta j(\beta,\mu;\Delta) &=& \lim_{L\to \infty} \frac{1}{L^3} \log(1-e^{-\beta(\epsilon_1(\bm{0})-\mu)}) + H(\beta,\mu)
\label{eq:eq_GFE}
\end{eqnarray}
with
\begin{eqnarray}
H(\beta,\mu) \equiv \int \frac{d^3 \bm{\rho}}{(2\pi)^2 \pi} \log (1-e^{-\beta (-2t \sum_{d=1}^3 \cos \rho_d - \mu)}).
\end{eqnarray}
Because $H(\beta,\mu)$ is independent of $\Delta$, $j(\beta,\mu;\Delta)$ depends on $\Delta$ only through the first term, which represents the contribution of the ground state. We consider the setting where $\rho$ is controlled through $\mu$ while $\beta$ is fixed. 

First, we consider the case $\Delta < 1/4$. Recalling (\ref{eq:limit 1}), we neglect the first term of (\ref{eq:eq_GFE}) for $\rho < G(\beta,-6t)$. We therefore obtain
\begin{eqnarray}
\beta j(\beta,\mu;\Delta < \frac{1}{4}) = H(\beta,\mu),
\end{eqnarray}
where $\mu < -6J$. For $\rho \geq G(\beta,-6J)$, $\mu$ is given by (\ref{eq:res_mu_SDLT0}). Substituting (\ref{eq:res_mu_SDLT0}) into (\ref{eq:eq_GFE}), we calculate the first term of (\ref{eq:eq_GFE}) as
\begin{eqnarray}
\lim_{L\to \infty} \frac{1}{L^3} \log(1-e^{-\beta (\epsilon_1(\bm{0})-\mu)}) &=& \lim_{L\to \infty} \frac{1}{L^3} \log(1-e^{-\frac{1}{L^3} \frac{1}{\rho-G(\beta,-6t)}}) \nonumber \\[3pt]
&=& 0.
\end{eqnarray}
We therefore obtain
\begin{eqnarray}
\beta j(\beta,\mu;\Delta < \frac{1}{4}) =  H(\beta,-6t)
\end{eqnarray}
in the thermodynamic limit. We thus conclude that the bulk free energy density is independent of $\Delta$. 

%%%%%%%%%%%%%%%%
%Note that one may study another setting which $\beta$ is controlled with $\rho$ fixed. We then consider the Helmholtz free energy density $f(\beta,\rho)$ defined as
%\begin{eqnarray}
%f(\beta,\rho) \equiv \max_{\mu} \{\mu \rho + j(\beta,\mu) \} .
%\label{eq:Legendre_f}
%\end{eqnarray}
%Because the maximizer is given by (\ref{eq:eq_mu_gen}), we obtain  
%\begin{eqnarray}
%\beta f(\beta,\rho;\Delta < \frac{1}{4}) = 
%\begin{cases}
%G^{-1}(\beta,\rho) \rho + H(\beta,G^{-1}(\beta,\rho)) & {\rm for} \ \beta<\beta_c ,\\[3pt]
%-6J \rho + H(\beta,- 6J) & {\rm for} \ \beta \geq \beta_c .
%\end{cases}
%\end{eqnarray}
%where $\beta_c$ is given by (\ref{eq:def_cri_temp_LD}). We also conclude that the Helmholtz free energy density is independent of $\Delta$.
%%%%%%%%%%%%%%%%

Second, we consider the case $\Delta \geq 1/4$. Through a similar calculation, for the case that $\rho$ is controlled with $\beta$ fixed, we obtain
\begin{align}
\beta j(\beta,\mu;\Delta \geq \frac{1}{4}) =
\begin{cases}
H(\beta,\mu) & {\rm for} \ \rho<G(\beta,\mu_{\ast}) ,\\[3pt]
H(\beta,\mu_{\ast}) & {\rm for} \ \rho\geq G(\beta,\mu_{\ast}).
\end{cases}
\end{align}
In the Bose--Einstein condensate, the free energy density explicitly depends on $\Delta$ because the condensation is localized at the surface layer with the enhanced coupling constant $t(1+\Delta)$.

%%%%%%%%%%%%%%%
%For the case that $\beta$ is controlled with $\rho$ fixed, we obtain
%\begin{align}
%\beta f(\beta,\rho;\Delta \geq \frac{1}{4}) = 
%\begin{cases}
%G^{-1}(\beta,\rho) \rho + H(\beta,G^{-1}(\beta,\rho)) & {\rm for} \ \beta<\beta_c  ,\\[3pt]
%\mu_{\ast} \rho + H(\beta,\mu_{\ast}) & {\rm for} \ \beta \geq \beta_c ,
%\end{cases}
%\end{align}
%where $\beta_c$ is given by (\ref{eq:def_cri_temp_HD}).
%%%%%%%%%%%%%%%%

\subsection{Non-analytic behavior}
We study the non-analyticity of the bulk free energy. In this section, we calculate the chemical potential $\mu$ as a function of $(\rho,\beta)$ by solving (\ref{eq:eq_mu_gen}) in $\mu$. In particular, we focus on the non-analytic behavior in $\beta$ near the critical point with $\rho$ fixed. 

\subsubsection{$\Delta<\frac{1}{4}$}
In the low-temperature region $\beta>\beta_c$, we obtained $\mu$ as a function of $(\rho,\beta)$ as shown in (\ref{eq:res_mu_SDLT0}) and (\ref{eq:res_mu_SDLT}). In the high-temperature region $\beta< \beta_c$, we expand $G(\beta,\mu)$ in $\mu$ around $\mu=-6t$. We define the new variable $m$ according to
\begin{eqnarray}
4 t m^2 \equiv \{(-6J) - \mu\}.
\end{eqnarray}
We then rewrite $G(\beta,\mu)$ in terms of $m$ as 
\begin{eqnarray}
G(\beta,\mu) = \int \frac{d^3 \bm{\rho}}{(2\pi)^2 \pi} \frac{1}{e^{4 \beta t (m^2 + \sum_{d=1}^3 \sin^2 \frac{\rho_d}{2})}-1}.
\end{eqnarray}
When $m$ is small, the dominant contribution to this integral arises from long-wavelength components. We explain this by considering the function
\begin{eqnarray}
\frac{\partial G(\beta,\mu)}{\partial \mu} = \int \frac{d^3 \bm{\rho}}{(2\pi)^2 \pi} \frac{\beta e^{4\beta t(m^2 + \sum_{d=1}^3 \sin^2\frac{\rho_d}{2})}}{\{e^{4 \beta t (m^2 + \sum_{d=1}^3 \sin^2 \frac{\rho_d}{2})}-1\}^2}.
\end{eqnarray}
We may rewrite this integral using the new variable $\bm{\rho} = m \bm{x}$ and focus on the contribution from small $m$. The integral is evaluated as
\begin{eqnarray}
\frac{\partial G(\beta,\mu)}{\partial \mu} \simeq m^{-1} \int d^3 \bm{x} \frac{1}{(1+|\bm{x}|^2)^2}.
\end{eqnarray}
This means that this function diverges as $m\to0$. Integrating this function in $\mu$, we obtain
\begin{eqnarray}
G(\beta,\mu) \simeq G(\beta,-6t) + A(\beta) ((-6t)-\mu)^{\frac{1}{2}},
\label{eq:expa_G_SD}
\end{eqnarray}
where $A(\beta)$ is a positive function depending on $\beta$. This expansion is valid when $\mu$ is near $-6t$. Substituting (\ref{eq:expa_G_SD}) into (\ref{eq:eq_mu_gen}), we obtain
\begin{eqnarray}
\rho \simeq G(\beta,-6t) + A(\beta) ((-6t)-\mu)^{\frac{1}{2}},
\end{eqnarray}
which leads to
\begin{eqnarray}
\mu - (-6t) \simeq - \frac{1}{A(\beta)^2} \biggl(\rho - G(\beta,-6t) \biggr)^2.
\label{eq:eq_mu_nearTc}
\end{eqnarray}
Because $A(\beta)$ and $G(\beta,-6t)$ do not exhibit any singularity, we expand the right-hand side of (\ref{eq:eq_mu_nearTc}) around $\beta=\beta_c$ as
\begin{eqnarray}
\mu - (-6t) \simeq -C \Big(\beta - \beta_c \Big)^2
\label{eq:mu_behavior_SDHT}
\end{eqnarray}
with
\begin{eqnarray}
C \equiv \Big(\frac{1}{A(\beta_c)} \frac{\partial G(\beta,-6t)}{\partial \beta} \Big|_{\beta=\beta_c}\Big)^2
\end{eqnarray}
for $\beta < \beta_c$, where we have used (\ref{eq:def_cri_temp_LD}).

From (\ref{eq:res_mu_SDLT}) and (\ref{eq:mu_behavior_SDHT}), we find that $\mu$ vanishes quadratically as $\beta \to \beta_c-0$ so that $\mu=\mu(\beta,\rho)$ has a discontinuous second derivative at $\beta_c$. This result is the same as that for $\Delta=0$. Recalling that the chemical potential $\mu$ is related to the Helmholtz free energy density through
\begin{eqnarray}
\mu = \biggl(\frac{\partial f}{\partial \rho} \biggr)_{\beta},
\label{eq:chemical potential and free energy}
\end{eqnarray}
we conclude that this Bose--Einstein condensation is the bulk critical phenomenon related to the non-analyticity of the bulk free energy density. 

\subsubsection{$\Delta \geq \frac{1}{4}$}
In the low-temperature region $\beta>\beta_c$, $\mu$ is given by (\ref{eq:res_mu_HDLT}). In the high-temperature region $\beta<\beta_c$, we expand $G(\beta,\mu)$ in $\mu$ around $\mu = \mu_{\ast}$. Unlike the case $\Delta<1/4$, the dominant contribution to the integral of $G(\beta,\mu)$ does not arise from the long-wavelength component and the derivative of $G(\beta,\mu)$ in $\mu$ does not diverge at $\mu= \mu_{\ast}$. We therefore have the expansion
\begin{eqnarray}
G(\beta,\mu) &\simeq& G(\beta,\mu_{\ast}) + \frac{\partial G(\beta,\mu)}{\partial \mu} \biggl|_{\mu=\mu_{\ast}} (\mu_{\ast} - \mu).
\label{eq:expa_G_HD}
\end{eqnarray}
Substituting (\ref{eq:expa_G_HD}) into (\ref{eq:eq_mu_gen}), we derive $\mu-\mu_{\ast}$ as
\begin{eqnarray}
\mu - \mu_{\ast} \simeq - \frac{1}{\frac{\partial G(\beta,\mu)}{\partial \mu} \biggl|_{\mu=\mu_{\ast}}} \Big(\rho - G(\beta,\mu_{\ast}) \Big).
\end{eqnarray}
Using (\ref{eq:def_cri_temp_HD}), we obtain
\begin{eqnarray}
\mu - \mu_{\ast} \simeq - C'(\beta_c-\beta)
\label{eq:mu_behavior_HDHT}
\end{eqnarray}
with
\begin{eqnarray}
C' \equiv \frac{\frac{\partial G(\beta,\mu_{\ast})}{\partial \beta}\Big|_{\beta=\beta_c}}{\frac{\partial G(\beta_c,\mu)}{\partial \mu} \Big|_{\mu=\mu_{\ast}}} > 0,
\end{eqnarray}
where $\beta<\beta_c$. The result implies that $\mu$ vanishes linearly as $\beta \to \beta_c-0$ so that $\mu=\mu(\beta,\rho)$ has a discontinuous first derivative at $\beta_c$. Because $f(\beta,\rho)$ has non-analyticity, we conclude that this Bose--Einstein condensation is described as the bulk phase transition. However the singularity type is different from that for $\Delta < 1/4$. It is noted that this singularity type is the same as that of the Bose--Einstein condensation in higher dimensions $d\geq4$.

\subsection{Non-analytic behavior of the specific heat}
We study the constant-volume specific heat. The internal energy $U(\beta,\rho)$ and internal energy density $u(\beta,\rho)$ are defined by
\begin{eqnarray}
U(\beta,\rho) \equiv  \sum_{n=1}^{\infty} \sum_{\bm{k}} \frac{\epsilon_n(\bm{k})}{e^{\beta (\epsilon_n(\bm{k})-\mu)}-1} \biggl |_{\mu=\mu(\beta,\rho)}
\end{eqnarray}
and 
\begin{eqnarray}
u(\beta,\rho) \equiv \lim_{L\to \infty} \frac{1}{L^3} U(\beta,\rho).
\end{eqnarray}
The internal energy density is rewritten as
%%
%\begin{eqnarray}
%u(\beta,\rho) = \lim_{L\to \infty} \frac{1}{L^3} \frac{\epsilon_1(\bm{0})}{e^{\beta(\epsilon_1(\bm{0})-\mu)}-1} + \int \frac{d^3 \bm{\rho}}{(2\pi)^2 \pi} \frac{-J \sum_{d=1}^3 \cos \rho_d}{e^{\beta(-J \sum_{d=1}^3 \cos \rho_d - \mu)}-1}
%\end{eqnarray}
%%
\begin{eqnarray}
u(\beta,\rho) = \rho \epsilon_1(\bm{0}) + I(\beta,\mu(\beta,\rho)),
\label{u: reexpression}
\end{eqnarray}
where $I(\beta,\mu)$ is defined by
\begin{eqnarray}
I(\beta,\mu) \equiv \int \frac{d^3 \bm{\rho}}{(2\pi)^2 \pi} \frac{-2t \sum_{d=1}^3 \cos \rho_d - \epsilon_1(\bm{0})}{e^{\beta(-2t \sum_{d=1}^3 \cos \rho_d - \mu)}-1}.
\end{eqnarray}
$I(\beta,\mu)$ is expanded in $\mu$ around $\mu=\epsilon_1(\bm{0})$ as
\begin{eqnarray}
I(\beta,\mu) \simeq I(\beta,\epsilon_1(\bm{0})) + \frac{\partial I(\beta,\mu)}{\partial \mu} \biggl |_{\mu=\epsilon_1(\bm{0})} (\mu- \epsilon_1(\bm{0}))
\label{eq:I expansion}
\end{eqnarray}
for $\mu \leq \epsilon_1(\bm{0})$. Substituting (\ref{eq:I expansion}) into (\ref{u: reexpression}), we obtain
\begin{eqnarray}
u(\beta,\rho) \simeq 
\begin{cases}
\epsilon_1(\bm{0}) \rho  + I(\beta,\epsilon_1(\bm{0})) +  \frac{\partial I(\beta,\mu)}{\partial \mu} \biggl |_{\epsilon_1(\bm{0})} (\mu(\beta,\rho) - \epsilon_1(\bm{0})) & {\rm for} \  \beta< \beta_c ,\\
\epsilon_1(\bm{0}) \rho + I(\beta,\epsilon_1(\bm{0})) & {\rm for} \ \beta \geq \beta_c.
\end{cases}
\label{eq:u expansion}
\end{eqnarray}
Note that (\ref{eq:u expansion}) holds regardless of $\Delta$. The constant-volume specific heat is given by the derivative of $u(\beta,\rho)$ in the temperature $1/\beta$. Clearly, the non-analyticity of the constant-volume specific heat originates from the term $\mu(\beta,\rho)-\epsilon_1(\bm{0})$. As shown in (\ref{eq:mu_behavior_SDHT}) and (\ref{eq:mu_behavior_HDHT}), the behavior of $\mu(\beta,\rho)-\epsilon_1(\bm{0})$ changes at $\Delta=1/4$. Therefore, the behavior of the constant-volume specific heat also changes at $\Delta=1/4$. When $\Delta \leq 1/4$, as is well known in the case $\Delta=0$, the constant-volume specific heat exhibits a cusp singularity at $\beta=\beta_c$. When $\Delta > 1/4$, the constant-volume specific heat has the discontinuous gap at $\beta = \beta_c$. The gap width depends on $\Delta$. 

%このように、自由エネルギーの非解析性の種類の違いは臨界点近傍での熱力学量の振る舞いの違いとして観測される。

\section{Surface free energy per unit area} 
\label{sec:Surface free energy per unit area}
We define an effective two-dimensional system by taking the partial trace over the degrees of freedom in the bulk. For convenience, we refer to this two-dimensional system as the effective surface system. This section studies the relation between the critical phenomenon in the effective surface system and the Bose--Einstein condensation in the bulk.

In this section, we consider the tight-binding Bose gas model on a cubic lattice
\begin{eqnarray}
\Lambda' = \{(i_1,i_2,j) \in \mathbb{Z}^3 \ |\ 1 \leq i_1 \leq M\ ,\ 1 \leq i_2 \leq M ,\ 1 \leq j \leq L \},
\label{eq:lattice1}
\end{eqnarray}
instead of (\ref{eq:lattice}). We explicitly express the $L$-dependence of each physical quantity by writing $L$ as the subscript. For example, the total free energy of the system is written as $J_L(\beta,\mu;\Delta)$.

\subsection{Surface free energy per unit area}
Let $J_{L}^0(\beta,\mu)$ denote the total free energy of the system with $\Delta=0$; i.e., 
\begin{eqnarray}
J_{L}^0(\beta,\mu) \equiv J_L(\beta,\mu;\Delta=0).
\end{eqnarray}
We define $J^s_L(\beta,\mu;\Delta)$ as
\begin{eqnarray}
J^s_L(\beta,\mu;\Delta) \equiv J_{L}(\beta,\mu;\Delta) - J_{L-1}^0(\beta,\mu).
\end{eqnarray}
We express $J^s_L(\beta,\mu)$ as
\begin{eqnarray}
e^{-\beta J^s_L(\beta,\mu)} = \int_{-\infty}^{\infty} \mathcal{D} (\bar{\psi}_1,\psi_1)  e^{- S^s_L[\bar{\psi}_1,\psi_1]},
\label{eq:Js}
\end{eqnarray}
where
\begin{eqnarray}
S^s_L[\bar{\psi}_1,\psi_1] \equiv \sum_{\bm{k},m} \bar{\psi}_{1,\bm{k},m} \mathcal{G}_L(\bm{k},\omega_m;\mu) \psi_{1,\bm{k},m},
\label{eq:Ss}
\end{eqnarray}
with
\begin{eqnarray}
\mathcal{G}_L(\bm{k},\omega_m;\mu) &\equiv& (A_{11}(\bm{k})-(\mu+i\omega_m)) -  t^2(B_{L-1}(\bm{k})- (\mu+i\omega_m) E_{L-1})^{-1}_{11}.
\label{eq:mathcal G}
\end{eqnarray}
See Appendix~\ref{sec:derivation of the effective surface system} for the derivation. $J^s_L(\beta,\mu)$ corresponds to the free energy of the effective surface system and $S^s_{L}[\bar{\psi}_1,\psi_1]$ corresponds to the action associated with the effective surface system. Then, we define $j^s(\beta,\mu;\Delta)$ as 
\begin{eqnarray}
j^s(\beta,\mu;\Delta) \equiv \lim_{M \to \infty} \frac{J^s_L(\beta,\mu;\Delta)}{M^2}.
\label{eq:the surface free energy per unit area}
\end{eqnarray}
We note that the total free energy $J_{L}(\beta,\mu;\Delta)$ depends on $M$ as the form
\begin{eqnarray}
J_{L}(\beta,\mu;\Delta) = L M^2 j(\beta,\mu;\Delta) + M^2 j^s(\beta,\mu;\Delta) + o(M^2)
\end{eqnarray}
in the thermodynamic limit~\cite{Binder2,Fisher and Caginalp,Caginalp and Fisher}. 

As shown in (\ref{eq:res_mu_SDLT}) and (\ref{eq:res_mu_HDLT}), the thermodynamic states below the critical point cannot be characterized by $\mu$. Therefore, we introduce the Helmholtz free energy as a function of $(\beta,\rho)$. Let $F_L(\beta,N;\Delta)$ be the total Helmholtz free energy. As with $J_L(\beta,\mu;\Delta)$, $F_L(\beta,N;\Delta)$ is expanded as
\begin{eqnarray}
F_{L}(\beta,N;\Delta) = L M^2 f(\beta,\rho;\Delta) + M^2 f^s(\beta,\rho;\Delta) + o(M^2)
\end{eqnarray}
in the thermodynamic limit, where $f(\beta,\mu;\Delta)$ and $f^s(\beta,\rho;\Delta)$ are defined as
\begin{eqnarray}
f(\beta,\rho;\Delta) \equiv \lim_{L,M \to \infty} \frac{F_{L}(\beta,N;\Delta)}{LM^2}
\end{eqnarray}
\begin{eqnarray}
f_s(\beta,\rho;\Delta) \equiv  \lim_{L,M \to \infty} \frac{F_{L}(\beta,N;\Delta) - LM^2 f(\beta,\rho;\Delta)}{M^2},
\end{eqnarray}
respectively. Here, $j(\beta,\mu;\Delta)$ and $f(\beta,\rho;\Delta)$ are connected by the Legendre transformation
\begin{eqnarray}
f(\beta,\rho;\Delta) = \max_{\mu} \{j(\beta,\mu;\Delta) + \mu \rho \}.
\label{eq:Legendre transformation}
\end{eqnarray}
We introduce $\bar{f}^s(\beta,\rho;\Delta)$ by extending the Legendre transformation as
\begin{eqnarray}
f(\beta,\rho;\Delta) + \frac{1}{L} \bar{f}_s(\beta,\rho;\Delta) =  \max_{\mu} \{j(\beta,\mu;\Delta) + \frac{1}{L}j^s(\beta,\mu;\Delta) + \mu \rho \}.
\label{eq:extended Legendre transformation}
\end{eqnarray}
Then, we assume that $f^s(\beta,\rho;\Delta)$ coincides with $\bar{f}^s(\beta,\rho;\Delta)$ in the thermodynamic limit. By noting that the argument of the maximum of (\ref{eq:Legendre transformation}) equals to that of (\ref{eq:extended Legendre transformation}) in the thermodynamic limit, we obtain
\begin{eqnarray}
f^s(\beta,\rho;\Delta) =  j^s(\beta,\mu(\beta,\rho);\Delta),
\label{eq:def fs}
\end{eqnarray}
where we have used the fact that the argument of the maximum of (\ref{eq:Legendre transformation}) is given by (\ref{eq:chemical potential and free energy}).

In the remainder of this section, we consider the non-analytic behavior of $f^s(\beta,\rho;\Delta)$ near the critical point. Focusing on the setting that $\beta$ is controlled with $\rho$ fixed, we study the non-analyticity of $f^s(\beta,\rho;\Delta)$ as a function of $\beta$. For simplicity, hereafter, we refer to $f^s(\beta,\rho;\Delta)$ as the surface free energy per unit area and the $\Delta$-dependence of $f^s(\beta,\rho;\Delta)$ is abbreviated.

Calculating the integral (\ref{eq:Js}), we obtain
\begin{eqnarray}
\beta J^s_L(\beta,\mu(\beta,\rho)) &=& \sum_{m=-\infty}^{\infty} \sum_{\bm{k}} \log \mathcal{G}_L(\bm{k},\omega_m;\mu(\beta,\rho)) \nonumber \\[3pt]
&=& \sum_{m(\neq0)} \sum_{\bm{k}} \log \mathcal{G}_L(\bm{k},\omega_m;\mu(\beta,\rho)) \nonumber \\[3pt]
&+& \sum_{\bm{k}} \log \Big((1+2\Delta)\omega(\bm{k})-\mu(\beta,\rho) + \sqrt{(\omega(\bm{k})-\mu(\beta,\rho))^2-(2t)^2} \Big),
\label{eq:Js_general}
\end{eqnarray}
where we have used (\ref{eq:G(0)}). We assume that the contribution of the component of $m\neq 0$ can be neglected as long as we focus on the non-analytic behavior of $f_s(\beta,\rho)$ near the critical point. We note that the validity of this assumption is confirmed for $\Delta \geq 1/4$ by straightforwardly computing the sum over Matsubara frequencies. From this assumption, we find that the non-analytic behavior of $f_s(\beta,\rho)$ originates from the last term of (\ref{eq:Js_general}). We then define 
\begin{eqnarray}
\beta F^s_{0L}(\beta,\rho) \equiv \sum_{|\bm{k}|<\Lambda} \log \Big((1+2\Delta)\omega(\bm{k})-\mu(\beta,\rho) + \sqrt{(\omega(\bm{k})-\mu(\beta,\rho))^2-(2t)^2} \Big),\quad
\label{eq:Js_0andlongwavelength}
\end{eqnarray}
where we have introduced the cutoff wavenumber $\Lambda$ because the contribution of the long-wavelength component is dominant in the non-analytic behavior of $f_s(\beta,\rho)$. Let $\Lambda$ be a sufficiently small but finite wave number. We then expand (\ref{eq:Js_0andlongwavelength}) around $\bm{k}=\bm{0}$ as
\begin{eqnarray}
\beta F_{0L}^s(\beta,\rho) &\simeq& \sum_{|\bm{k}|<\Lambda} \log \Big(\{-4t(1+2\Delta) - \mu(\beta,\rho)\}+t(1+2\Delta)\bm{k}^2 \nonumber \\[3pt]
&+& \sqrt{(4t+\mu(\beta,\rho))^2 -(2t)^2 - 2t(4t+\mu(\beta,\rho))\bm{k}^2} \Big),
\label{eq:Js_expand}
\end{eqnarray}
where we have used (\ref{eq:Definition Omega(k)}). We define $f_0^s(\beta,\rho)$ as
\begin{eqnarray}
f^s_0(\beta,\rho) \equiv \lim_{L\to\infty} \frac{F^s_{0L}(\beta,\rho)}{M^2}.
\end{eqnarray}
As long as we focus on the non-analyticity, we have only to study $f_0^s(\beta,\rho)$ instead of $f^s(\beta,\rho)$.

Here, we note that the surface free energy $f^s_0(\beta,\rho)$ is separated into two parts. We express the argument of the logarithm of (\ref{eq:Js_expand}) as
\begin{eqnarray}
I(\beta,\rho) &\equiv& I_1(\beta,\rho) + I_2(\beta,\rho),
\end{eqnarray}
with
\begin{eqnarray}
I_1(\beta,\rho) =  \{-4t(1+2\Delta) - \mu(\beta,\rho)\}+t(1+2\Delta)\bm{k}^2 ,
\label{eq:I1I1I1}
\end{eqnarray}
\begin{eqnarray}
I_2(\beta,\rho) = \sqrt{(4t+\mu(\beta,\rho))^2 -(2t)^2 - 2t(4t+\mu(\beta,\rho))\bm{k}^2}.
\end{eqnarray}
$I_1(\beta,\rho)$ and $I_2(\beta,\rho)$ come from the first and second terms of $\mathcal{G}_L(\bm{k},\omega_m;\mu)$ (\ref{eq:mathcal G}), respectively. $\mathcal{G}_L(\bm{k},\omega_m;\mu)$ is obtained by taking the partial trace of (\ref{eq:grand canonical free energy}) over the degrees of freedom in the bulk (See Appendix~\ref{sec:derivation of the effective surface system}). Through this calculation, we find that the first term of $\mathcal{G}_L(\bm{k},\omega_m;\mu)$ represents direct coupling of the degrees of freedom at the surface, and the second term the indirect coupling of the degrees of freedom at the surface through the degrees of freedom in the bulk. Therefore, $I_1(\beta,\rho)$ represents the effect of the pure two-dimensional system with the hopping rate $(1+\Delta)t$ and the chemical potential $\mu(\beta,\rho)$. $I_2(\beta,\rho)$ represents the effect of the interaction through the bulk. Here, we note that $I_1(\beta,\rho)$ is also affected by the bulk through the chemical potential. This connection stems from the conservation of the particle number of the total system. %$f_0^s(\beta,\rho)$ consists of these two parts $I_1(\beta,\rho)$ and $I_2(\beta,\rho)$.
We below demonstrate that the non-analytic behavior of $f_0^s(\beta,\rho)$ at $\beta=\beta_c$ is understood by the competition between $I_1(\beta,\rho)$ and $I_2(\beta,\rho)$

We define the constant-volume specific heat $c_v^s$ of the effective surface system as
\begin{eqnarray}
c_v^s \equiv -T \Big(\frac{\partial^2 f_s(\beta,\rho)}{\partial T^2} \Big)_{\rho} = - \frac{\beta^2}{k_B} \Big(\frac{\partial^2}{\partial \beta^2}\beta f_s(\beta,\rho) \Big)_{\rho}.
\label{eq:def_cvs}
\end{eqnarray}
To study the non-analytic behavior of $f^s(\beta,\rho)$, we calculate the non-analytic behavior of $c_v^s$ in $\beta$ near the critical point with $\rho$ fixed.

\subsection{Singularity type for the case $\Delta<\frac{1}{4}$}
Near the critical point $\beta=\beta_c$, $\mu(\beta,\rho)$ is given by (\ref{eq:res_mu_SDLT}) and (\ref{eq:mu_behavior_SDHT}). Substituting (\ref{eq:res_mu_SDLT}) and (\ref{eq:mu_behavior_SDHT}) into (\ref{eq:Js_expand}), we express $F^s_{0L}(\beta,\rho)$ as
\begin{eqnarray}
\beta F^s_{0L}(\beta,\rho) &\simeq& \sum_{|\bm{k}|<\Lambda} \log \Big(I^{<}_1(\beta,\rho) + I^{<}_2(\beta,\rho) \Big),
\label{eq:Fs0L _HT}
\end{eqnarray}
where $I^{<}_1(\beta,\rho)$ and $I^{<}_2(\beta,\rho)$ are given by
\begin{eqnarray}
I^{<}_1(\beta,\rho) = 
\begin{cases}
2t(1+4\Delta) +C(\beta-\beta_c)^2+t(1+2\Delta)\bm{k}^2 &{\rm for} \  \beta<\beta_c,  \\[3pt]
2t(1+4\Delta) +t(1+2\Delta)\bm{k}^2 & {\rm for} \  \beta>\beta_c,
\end{cases}
\label{eq:I1< beh}
\end{eqnarray}
and
\begin{eqnarray}
I^{<}_2(\beta,\rho) = 
\begin{cases}
\sqrt{4tC(\beta-\beta_c)^2 + C^2(\beta-\beta_c)^4 + 2t(2t+C(\beta-\beta_c)^2)\bm{k}^2}&{\rm for} \  \beta<\beta_c, \\[3pt]
\sqrt{4t^2\bm{k}^2} & {\rm for} \ \beta>\beta_c.
\end{cases}
\label{eq:I2< beh}
\end{eqnarray}
We focus on the behavior of $I^{<}_2(\beta,\rho)$ at the limit $|\bm{k}|\to 0$. $|\bm{k}|$-dependence of $I^{<}_2(\beta,\rho)$ changes from $|\bm{k}|^2$ to $|\bm{k}|$ at $\beta=\beta_c$. The linearity of $I^{<}_2(\beta,\rho)$ in $|\bm{k}|$ for $\beta>\beta_c$ means that there are long-range interaction between the degrees of freedom at surface. By recalling that $I^{<}_2(\beta,\rho)$ represents the effect of the interaction through the bulk, we conclude that below the bulk critical point, the ordered bulk induces the long-range interaction in the surface free energy. Such long-range interaction leads to the non-analyticity of $f^s(\beta,\rho)$.

We define $u^s_0(\beta,\rho)$ as
\begin{eqnarray}
u^s_0(\beta,\rho) \equiv - \Big( \frac{\partial}{\partial \beta} \beta f^s_0(\beta,\rho)\Big)_{\rho}.
\label{eq:us0}
\end{eqnarray}
We exress the leading term in $u^s_0(\beta,\rho)$ as $|\beta-\beta_c| \to 0$. We first consider the case $\beta < \beta_c$. Substituting (\ref{eq:Fs0L _HT}) into (\ref{eq:us0}), we obtain
\begin{eqnarray}
u^s_0(\beta,\rho) = - \frac{1}{M^2} \sum_{|\bm{k}|<\Lambda} \frac{1}{I^{<}_1(\beta,\rho) + I^{<}_2(\beta,\rho)} \Big(\frac{\partial I^{<}_1(\beta,\rho)}{\partial \beta} + \frac{\partial I^{<}_2(\beta,\rho)}{\partial \beta} \Big).
\label{eq:us0_original}
\end{eqnarray}
Here, using (\ref{eq:I1< beh}) and (\ref{eq:I2< beh}), we calculate the behavior of each part of (\ref{eq:us0_original}) in the limits $|\beta-\beta_c| \to 0$ and $\bm{k} \to \bm{0}$ as
\begin{eqnarray}
\frac{1}{I^{<}_1(\beta,\rho) + I^{<}_2(\beta,\rho)} \simeq \frac{1}{2t(1+4\Delta)},
\end{eqnarray}
\begin{eqnarray}
\frac{\partial I^{<}_1(\beta,\rho)}{\partial \beta} \simeq 2C(\beta-\beta_c),
\end{eqnarray}
\begin{eqnarray}
\frac{\partial I^{<}_2(\beta,\rho)}{\partial \beta} \simeq (\beta-\beta_c) \frac{2tC}{\sqrt{tC(\beta-\beta_c)^2 + t^2\bm{k}^2}}.
\end{eqnarray}
Therefore, we calculate $u^s_0(\beta,\rho)$ in the limit $M \to \infty$ as
\begin{eqnarray}
u^s_0(\beta,\rho) &=& (\beta_c-\beta) \frac{1}{M^2} \sum_{|\bm{k}|<\Lambda}  \frac{C\sqrt{tC(\beta-\beta_c)^2 + t^2\bm{k}^2} + C}{t(1+4\Delta)}\frac{1}{\sqrt{tC(\beta-\beta_c)^2 + t^2\bm{k}^2}} \nonumber \\[3pt]
&\simeq&  (\beta_c-\beta) \frac{C}{t(1+4\Delta)} \int_{|\bm{k}|<\Lambda} d^2\bm{k} \frac{1}{\sqrt{4tC(\beta-\beta_c)^2 + 4t^2\bm{k}^2}} .
\label{eq:us0_calcu}
\end{eqnarray}
Since the integral in (\ref{eq:us0_calcu}) converges at $\beta \to \beta_c$, we obtain $\beta-\beta_c$-dependence of $u^s_0(\beta,\rho)$ as
\begin{eqnarray}
u^s_0(\beta,\rho)&\sim& (\beta_c-\beta).
\label{eq:u0s_behavior}
\end{eqnarray}
We note that $|\bm{k}|$-dependence of $I_2^{<}(\beta,\rho)$ leads to the fact that the integral in (\ref{eq:us0_calcu}) converges at $\beta \to \beta_c$.

We next consider the case $\beta \geq \beta_c$. In the similar calculation, we obtain
\begin{eqnarray}
u^s_0(\beta,\rho) = 0.
\label{eq:u0s_behavior_low}
\end{eqnarray}
The singular behavior of the constant-volume specific heat originates from the derivative of $u^s_0(\beta,\rho)$ in the temperature $1/\beta$. From (\ref{eq:u0s_behavior}) and (\ref{eq:u0s_behavior_low}), we find that the constant-volume specific heat of the effective surface system exhibits a discontinuity at $\beta=\beta_c$. By recalling that the constant-volume specific heat of the bulk exhibits a cusp singularity at $\beta = \beta_c$, we find that the singularity type of $f^s(\beta,\rho)$ is different from that of $f(\beta,\rho)$, which is expressed in terms of  the singular part of the free energy as
\begin{eqnarray}
f(\beta,\rho) \sim (\beta_c-\beta)^3,
\label{eq:exponent f}
\end{eqnarray}
and
\begin{eqnarray}
f^s(\beta,\rho) \sim (\beta_c-\beta)^2
\label{eq:exponent fs}
\end{eqnarray}
for $\beta<\beta_c$. The difference of the exponents in (\ref{eq:exponent f}) and (\ref{eq:exponent fs}) is equal to one. This property was observed for many models such as Ising model and $\phi^4$ model by means of the renormalization group method~\cite{Diehl2}. We note that the singularity of $f^s(\beta,\rho)$ comes from $|\bm{k}|$-dependence of $I^{<}_2(\beta,\rho)$. This means that the critical phenomena of the effective surface system are induced by the ordered bulk.

\subsection{Singularity type for the case $\Delta\geq \frac{1}{4}$}
Near the critical point $\beta=\beta_c$, the behavior of $\mu(\beta,\rho)$ is given by (\ref{eq:res_mu_HDLT}) and (\ref{eq:mu_behavior_HDHT}). Using these results, we express $F^s_{0L}(\beta,\rho)$ as
\begin{eqnarray}
\beta F^s_{0L}(\beta,\rho) %&\simeq& \sum_{\bm{k}<\Lambda} \log \Big(\{J\frac{1+16\Delta^2}{4\Delta}-8J\Delta\} -C'(\beta-\beta_c)+ J(1+2\Delta)\bm{k}^2 \nonumber \\[3pt]
%&+& \sqrt{(J\frac{1+16\Delta^2}{4\Delta})^2 - J\frac{1+16\Delta^2}{2\Delta}C'(\beta-\beta_c) + C'^2(\beta-\beta_c)^2 - (2J)^2 +J^2\frac{1+16\Delta^2}{2\Delta}\bm{k}^2}\Big) \nonumber \\[3pt]
&\simeq&  \sum_{\bm{k}<\Lambda} \log \Big(I^{>}_1(\beta,\rho) + I^{>}_2(\beta,\rho)  \Big),
\label{eq:Fs0L last0}
\end{eqnarray}
where $I^{>}_1(\beta,\rho)$ and $I^{>}_2(\beta,\rho)$ are given by
\begin{eqnarray}
I^{>}_1(\beta,\rho) = 
\begin{cases}
(c-8\Delta)t + C'(\beta_c-\beta) + t(1+2\Delta) \bm{k}^2 &{\rm for} \  \beta<\beta_c, \\[3pt]
(c-8\Delta)t + t(1+2\Delta) \bm{k}^2 & {\rm for} \  \beta>\beta_c,
\end{cases}
\end{eqnarray}
and
\begin{eqnarray}
I^{>}_2(\beta,\rho) =  
\begin{cases}
\sqrt{\Big(ct+C'(\beta_c-\beta)\Big)^2-4t^2-2t\Big(-ct -C'(\beta_c-\beta)\Big)\bm{k}^2}&{\rm for} \  \beta<\beta_c , \\[3pt]
\sqrt{\Big(c^2-4\Big)t^2+2ct^2\bm{k}^2} & {\rm for} \ \beta>\beta_c 
\end{cases}
\end{eqnarray}
with
\begin{eqnarray}
c \equiv \frac{1+16\Delta^2}{4\Delta}.
\end{eqnarray}
The singularity of the surface free energy comes from combining $I^{>}_1(\beta,\rho)$ and $I^{>}_2(\beta,\rho)$. In order to show this, we expand $I^{>}_2(\beta,\rho)$ in $\bm{k}$ as
\begin{eqnarray}
I^{>}_2(\beta,\rho) = \sqrt{\Big(ct+C'(\beta_c-\beta)\Big)^2-4t^2} - \frac{t\Big(-ct -C'(\beta_c-\beta)\Big)}{\sqrt{\Big(ct+C'(\beta_c-\beta)\Big)^2-4t^2}} \bm{k}^2 + O(\bm{k}^4)
\label{eq:I>2 intermediate}
\end{eqnarray}
for $\beta<\beta_c$. By noting 
\begin{eqnarray}
\sqrt{c^2-4} = 8\Delta-c,
\end{eqnarray}
we expand (\ref{eq:I>2 intermediate}) in $\beta_c-\beta$ as
\begin{eqnarray}
I^{>}_2(\beta,\rho) &=& (8\Delta-c)t + \frac{cC'}{8\Delta-c}(\beta_c-\beta) \nonumber \\[3pt]
&+& \frac{ct}{8\Delta-c} \bm{k}^2 - \frac{4C'}{(8\Delta -c)^3} (\beta_c-\beta)\bm{k}^2 + O(\bm{k}^4,(\beta_c-\beta)^2)
\end{eqnarray}
for $\beta<\beta_c$. By combining $I^{>}_1(\beta,\rho)$ and $I^{>}_2(\beta,\rho)$, we obtain
\begin{eqnarray}
I^{>}_1(\beta,\rho) + I^{>}_2(\beta,\rho) = A_1(\beta_c-\beta) + A_2(\beta_c-\beta) \bm{k}^2 + O(\bm{k}^4)
\label{eq:158}
\end{eqnarray}
with
\begin{eqnarray}
A_1(x) = \frac{8\Delta}{8\Delta-c}C' x + O(x^2),
\label{eq:beta dependence of A1}
\end{eqnarray}
\begin{eqnarray}
A_2(x) = t(1+2\Delta) + \frac{ct}{8\Delta-c} - \frac{4C'}{(8\Delta -c)^3} x + O(x^2).
\label{eq:beta dependence of A2}
\end{eqnarray}
We immediately confirm $A_1(0)=0$ and $A_2(0)\neq0$. Substituting (\ref{eq:158}) into (\ref{eq:Fs0L last0}) and calculating the sum of (\ref{eq:Fs0L last0}) in the limit $M \to \infty$, we find that $F^s_{0L}$ is $O(M^2 \log M)$. The surface free energy per unit area defined by (\ref{eq:the surface free energy per unit area}) therefore exhibits logarithmic divergence in the limit $M \to \infty$. From (\ref{eq:Fs0L last0}), (\ref{eq:158}), ({\ref{eq:beta dependence of A1}) and ({\ref{eq:beta dependence of A2}), we find that $F^s_0(\beta,\rho)$ is the same as that of the two-dimensional Gauss model. We therefore obtain the behavior of the constant-volume specific heat as
\begin{eqnarray}
c_v^s \sim \frac{1}{|\beta_c-\beta|}
\end{eqnarray}
for $\beta<\beta_c$. The singularity type of the constant-volume specific heat of the effective surface system is different from that defined from the bulk free energy. This means that the type of non-analyticity of $f^s(\beta,\rho)$ is different from that of $f(\beta,\rho)$. We note that the singular behavior of the Gauss model comes from the $x$-dependence of $A_1(x)$ as
\begin{eqnarray}
A_1(x) \sim x \ {\rm for} \ x \to 0.
\end{eqnarray}
This behavior results from combining $I^{>}_1(\beta,\rho)$ and $I^{>}_2(\beta,\rho)$. Therefore, we confirm that the critical phenomena in the effective surface system are induced by connecting the ordered bulk and the surface.

\section{Conclusion and discussion}
%この論文で扱ったこと
In this paper, we investigated the Bose--Einstein condensation in the tight-binding model with the hopping rate enhanced only on a surface. Regardless of the strength of the enhanced hopping rate $t_s$, this model exhibits Bose--Einstein condensation. However, we found that two different critical behaviors occur depending on the case $t_s < 5/4$ or $t_s \geq 5/4$. We analyzed the critical behaviors as the bulk critical phenomenon and the surface critical phenomenon.

%$J_s/J<5/4$の時
In the case $t_s/t<5/4$, our model exhibits the same critical behavior as in the unenhanced case $t_s=t$. At the criticality of the bulk, the singular parts of the bulk and surface free energy are expressed in the form
\begin{eqnarray}
f(\beta,\rho) \sim (\beta_c-\beta)^3,
\end{eqnarray}
and
\begin{eqnarray}
f^s(\beta,\rho) \sim (\beta_c-\beta)^2
\end{eqnarray}
for $\beta<\beta_c$.

%$J_s/J>5/4$の時
In the case $t_s/t \geq 5/4$, unique and pathological Bose--Einstein condensation occurs. The singularity type of the bulk free energy is the same as that of the Bose--Einstein condensation with $t_s=t$ in higher dimensions $d \geq 4$. Meanwhile, the singularity type of the surface free energy is the same as that of the two-dimensional Gauss model. In the Bose--Einstein condensate, $O(L^3)$ particles are spatially localized in the surface layer. As a result, the bulk free energy density explicitly depends on $t_s$ defined only at the surface. Furthermore, the surface free energy per unit area exhibits logarithmic divergence in the thermodynamic limit. The chemical potential $\mu$ also explicitly depends on $t_s$ for the Bose--Einstein condensate.

%surface free energyの解析
We also analyzed these two Bose--Einstein condensation in terms of the surface free energy. By exactly calculating the surface free energy, we showed that the surface free energy is divided into two parts. The first part corresponds to the pure two-dimensional system with the hopping rate $t_s$ and the chemical potential $\mu(\beta,\rho)$. The second part corresponds to the interaction between the degree of freedom at the surface through the bulk. The singularity type of the surface free energy is determined by the competition between these parts. In the case $t_s/t < 5/4$, the origin of the singularity of the surface free energy is the long-range interaction through the ordered bulk. Such phase transition is often referred to as the "ordinary transition"~\cite{Binder1}. In the case $t_s/t \geq 5/4$, it arises from connecting the enhanced hopping effects at the surface and the divergence of the correlation function in bulk. Such phase transition is often referred to as the "special transition"~\cite{Binder1}.

%今後の課題1 - 相互作用を取り入れたらどうなるか？
We note that the two-dimensional localization of $O(L^3)$ particles is possible because our model is a collection of free bosons. If we take a repulsive interaction between the bosons into account, such spatial localization can never occur. Let us consider the Bose--Einstein condensation in the experiments of dilute Bose gas that the interaction between bosons is tuned sufficiently small. When the system is sufficiently dilute that most particles composing the system can be gathered in the two-dimensional surface, we expect that the critical exponents are given by that for the ideal gas. On the other hand, when we focus on more dense systems, we must pay attention to the repulsive interaction between the bosons. It remains unclear how the results for the ideal gas are modified by the repulsive interaction and what we observe in real experimental systems.

%今後の課題2 - 対称性だけではダメかもしれないこと
If we consider the surface critical phenomena by focusing only on the symmetry of the system, we find that the surface critical phenomena of interacting bosons are the same as the XY model. However, we conjecture that the surface critical phenomena cannot be determined only by the symmetry. As shown in this paper, $f^s(\beta,\rho)$ is determined by the competition between the bulk and the surface (especially in the special transition). We consider that the coupling between the bulk and the surface may depend on the elements other than the symmetry. A possible element is, for example, the conservation law. In the collection of the bosons, the number of particles is conserved, which is in contrast to the XY model. Thus, the next step is to study the difference between the interacting bosons and XY model.

%念のため、
We comment that even in the ideal Bose gas, the conservation law plays a non-trivial role in coupling between the bulk and the surface. By noting that the direct coupling of the degrees of freedom at the surface in the surface free energy $f^s(\beta,\rho)$, (\ref{eq:I1I1I1}), depends on the chemical potential $\mu(\beta,\rho)$, we find that the surface critical phenomena are strongly affected by the bulk through the chemical potential. Since the chemical potential is associated with the particle number of the total system, we confirm that the conservation law plays the important role in the surface critical phenomena.

%%修正の必要あり
%今後への課題0 - アンサンブルの選択について
Let us return to the definition of the surface free energy per unit area (\ref{eq:def fs}). We assumed that in the thermodynamics limit, $f^s(\beta,\rho)$ coincides with $\bar{f}^s(\beta,\rho)$. Here, $\bar{f}^s(\beta,\rho)$ is defined by extending the Legendre transformation (\ref{eq:extended Legendre transformation}). However, it remains to be elucidated whether the equivalence of ensembles holds even on the level of the $O(L^{-1})$ correction term of the free energy. In general, the expectation of the local quantity depends on the choice of the ensemble. In the Bose--Einstein condensate, for example, the second cumulant of the particle number occupying the ground state depends on the choice of ensemble, although the fraction of the ground state atoms is independent of the choice of ensemble~\cite{Mewes,Groot,Navez}. Therefore, it is required that we more carefully discuss the difference between $f^s(\beta,\rho)$ and $\bar{f}^s(\beta,\rho)$.

%今後への課題2 - 非平衡にしたらどうなるか？
We finally remark on the possibility of the surface long-range order in nonequilibrium steady states of systems with conservative dynamics. In particular, it is interesting to consider the case that these systems are subjected to an external forcing parallel to the surface. It is known that the spatial correlations of fluctuations of conserved quantities generally decay via a power law~\cite{Spohn,Dorfman}. We expect that this spatially long-range correlation leads to the long-range interaction between the degrees of freedom at the surface and induces the surface long-range order.

\begin{acknowledgements}
The authors would like to thank H.~Tasaki, K.~Adachi and Y.~Minami for helpful comments. The present study was supported by KAKENHI (Nos. 25103002 and 17H01148).
\end{acknowledgements}

\appendix\normalsize
\renewcommand{\theequation}{\Alph{section}.\arabic{equation}}
\setcounter{equation}{0}
\makeatletter
  \def\@seccntformat#1{%
    \@nameuse{@seccnt@prefix@#1}%
    \@nameuse{the#1}%
    \@nameuse{@seccnt@postfix@#1}%
    \@nameuse{@seccnt@afterskip@#1}}
  \def\@seccnt@prefix@section{Appendix }
  \def\@seccnt@postfix@section{:}
  \def\@seccnt@afterskip@section{\ }
  \def\@seccnt@afterskip@subsection{\ }
\makeatother

\section{energy eigenvalues for the case $\omega(\bm{k}) \geq 0$} \label{appsec:A}
First, we consider the energy eigenvalues for the case $\omega(\bm{k})=0$. Substituting $\omega(\bm{k})=0$ into (\ref{eq:Determinant A}), we obtain
\begin{eqnarray}
f(z;\bm{k}) = \det \Big[ B_{L} (\bm{k}) - z E_L \Big] .
\end{eqnarray}
This result immediately leads to
\begin{eqnarray}
\epsilon_n(\bm{k}) = \epsilon_n^0(\bm{k}).
\label{eq:eps_n=eps_n0}
\end{eqnarray}

Second, we consider the energy eigenvalues for the case $\omega(\bm{k})>0$. We define $\bm{k}'$ as
\begin{eqnarray}
\omega(\bm{k}') = - \omega(\bm{k}).
\end{eqnarray}
Noting
\begin{eqnarray}
\det \Big[ B_{L} (\bm{k}') - z E_L \Big] &=& \prod_{n=1}^{L} \Big(-z - \omega(\bm{k}) - 2t \cos(\frac{n\pi}{L+1}) \Big) \nonumber \\[3pt]
&=& (-1)^L \prod_{n=1}^{L} \Big(- (-z) + \omega(\bm{k}) - 2t \cos(\frac{n\pi}{L+1}) \Big) \nonumber \\[3pt]
&=& (-1)^L \det \Big[ B_{L} (\bm{k}) - (-z) E_L \Big],
\end{eqnarray}
we obtain
\begin{eqnarray}
f(-z;\bm{k}') = (-1)^L f(z;\bm{k}).
\end{eqnarray}
Therefore we associate $\epsilon_n(\bm{k})$ with $\omega(\bm{k})<0$ to $\epsilon_n(\bm{k}')$ with $\omega(\bm{k'}) = - \omega(\bm{k})>0$ as 
\begin{eqnarray}
\epsilon_n(\bm{k}) = - \epsilon_{L-(n-1)}(\bm{k}').
\label{eq:relation k and k'}
\end{eqnarray}

\section{Solution of (\ref{eq:Characteristic Equation A After Deformation})}
\label{sec:solution of (42)}
We derive the solution $z$ of (\ref{eq:Characteristic Equation A After Deformation}). In particular, because we have considered only the range $z<\omega(\bm{k}) - 2t$ to transform the characteristic equation (\ref{eq:Characteristic Equation A}) into (\ref{eq:Characteristic Equation A After Deformation}), we focus on the solution $z$ within $z<\omega(\bm{k}) - 2t$.

We define the function
\begin{eqnarray}
g(x) \equiv \frac{1}{2} \Big(x + \Delta \omega(\bm{k}) \Big) \Big( \sqrt{x^2- 4t^2} + x \Big) - t^2 .
\end{eqnarray}
Using this function $g(x)$, we rewrite (\ref{eq:Characteristic Equation A After Deformation}) as
\begin{eqnarray}
g(-z+\omega(\bm{k})) = 0 .
\label{eq:Characteristic Equation A Using g}
\end{eqnarray}
Because $g(x)$ is the monotonically increasing function of $x$ for $x>2t$, (\ref{eq:Characteristic Equation A Using g}) has a solution for $z<\omega(\bm{k})-2t$ when
\begin{eqnarray}
g(2t) = t \Big(2t + \Delta \omega(\bm{k}) \Big) - t^2 < 0 .
\end{eqnarray}
Simplifying this condition, we obtain
\begin{eqnarray}
\Delta \omega(\bm{k}) < - t .
\label{eq:k range solution exist appendix}
\end{eqnarray}
Therefore (\ref{eq:Characteristic Equation A After Deformation}) has the solution for wave numbers satisfying (\ref{eq:k range solution exist appendix}). Solving (\ref{eq:Characteristic Equation A After Deformation}), we obtain the solution 
\begin{eqnarray}
z = \frac{t^2+\Delta^{2}\omega(\bm{k})^2}{\Delta \omega(\bm{k})} + \omega(\bm{k}) .
\end{eqnarray}
Meanwhile for wave numbers
\begin{eqnarray}
\Delta \omega(\bm{k}) \geq - t,
\label{eq:k range solution not exist appendix}
\end{eqnarray}
(\ref{eq:Characteristic Equation A After Deformation}) does not have the solution for $z<\omega(\bm{k})-2t$.

\section{Proof that (\ref{eq:eq_EE}) and (\ref{eq:eq_U}) satisfy the eigenvalue equation (\ref{eq:Eigenvalue Equation vector})}
\label{sec:chech of vn}
We show that (\ref{eq:eq_EE}) and (\ref{eq:eq_U}) satisfy the eigenvalue equation (\ref{eq:Eigenvalue Equation vector}). We substitute (\ref{eq:eq_U}) into the left-hand side of (\ref{eq:Eigenvalue Equation vector}). 
\begin{eqnarray}
\Big(A(\bm{k}) \bm{v}_n(\bm{k}) \Big)_1 &=& (1+\Delta) \omega(\bm{k}) v^1_n(\bm{k}) - t v^2_n(\bm{k}) \nonumber \\[3pt]
&=& \frac{v_n^1(\bm{k})}{\sin \theta_n(\bm{k})} \Big((1+\Delta) \omega(\bm{k}) \Big\{\sin(\theta_n(\bm{k})) \Big\} - t \Big\{\sin(2\theta_n(\bm{k})) + \frac{\Delta \omega(\bm{k})}{t} \sin(\theta_n(\bm{k}))\Big\} \Big) \nonumber \\[3pt]
&=& \frac{v_n^1(\bm{k})}{\sin \theta_n(\bm{k})} \Big(\omega(\bm{k}) \sin(\theta_n(\bm{k})) - t \sin(2\theta_n(\bm{k})) \Big) \nonumber \\[3pt]
&=& \epsilon_n(\bm{k}) v_n^1(\bm{k})
\label{eq:C13}
\end{eqnarray}
Before proceeding to the calculation for $2 \leq j \leq L$, we note that (\ref{eq:eq_THETA}) leads to $v_n^{L+1}(\bm{k})=0$. Taking this property into account, we obtain
\begin{eqnarray}
v^{j-1}_n(\bm{k}) + v^{j+1}_n(\bm{k}) &=& \frac{v_n^1(\bm{k})}{\sin \theta_n(\bm{k})}\Big\{\sin((j-1)\theta_n(\bm{k})) + \frac{\Delta \omega(\bm{k})}{t} \sin((j-2)\theta_n(\bm{k}))\Big\} \nonumber \\[3pt]
&+&\frac{v_n^1(\bm{k})}{\sin \theta_n(\bm{k})}\Big\{\sin((j+1)\theta_n(\bm{k})) + \frac{\Delta \omega(\bm{k})}{t} \sin(j\theta_n(\bm{k}))\Big\} \nonumber \\[3pt]
&=&  \frac{v_n^1(\bm{k})}{\sin \theta_n(\bm{k})} \times 2 \cos (\theta_n(\bm{k})) \Big(\sin (j\theta_n(\bm{k})) + \frac{\Delta \omega(\bm{k})}{t}\sin ((j-1)\theta_n(\bm{k})) \Big),\nonumber\\
\label{eq:C14}
\end{eqnarray}
for $2 \leq j \leq L$. Using (\ref{eq:eq_U}) and (\ref{eq:C14}), the left-hand side of (\ref{eq:Eigenvalue Equation vector}) is rewritten as
\begin{eqnarray}
\Big(A(\bm{k}) \bm{v}_n(\bm{k}) \Big)_j &=& - t v^{j-1}_n(\bm{k}) + \omega(\bm{k}) v^{j}_n(\bm{k}) - t v^{j+1}_n(\bm{k}) \nonumber \\[3pt]
&=& \frac{v_n^1(\bm{k})}{\sin \theta_n(\bm{k})} \bigg[ - 2 t\cos(\theta_n(\bm{k})) \Big\{\sin (j\theta_n(\bm{k})) + \frac{\Delta \omega(\bm{k})}{t}\sin ((j-1)\theta_n(\bm{k})) \Big\}  \nonumber \\[3pt]
&+& \omega(\bm{k}) \Big\{\sin(j\theta_n(\bm{k})) + \frac{\Delta \omega(\bm{k})}{t} \sin((j-1)\theta_n(\bm{k})) \Big\} \bigg] \nonumber \\[3pt]
&=& \epsilon_n(\bm{k}) v_n^j(\bm{k})
\label{eq:C15}
\end{eqnarray}
where $2 \leq j \leq L$. Equations (\ref{eq:C13}) and (\ref{eq:C15}) conclude that (\ref{eq:eq_EE}) and (\ref{eq:eq_U}) satisfy the eigenvalue equation (\ref{eq:Eigenvalue Equation vector}).

\section{Derivation of some formulae by using path integral expression} \label{sec:PIaL}
We use the path integral method to derive some formulae in this paper. We shall summarize details of the calculation.

\subsection{Preliminary}
Introducing the path integral expression, the partition function of the grand canonical ensemble is written as
\begin{eqnarray}
e^{- \beta J(\beta,\mu)} &=& \sum_{n} <\bm{n}|\sum_{\bm{k}} e^{-\beta(\hat{H}-\mu \hat{N})}|\bm{n}> \nonumber \\[3pt]
&=& \int_{-\infty}^{\infty} \mathcal{D} (\bar{\psi},\psi)  e^{- S[\bar{\psi},\psi]}
\label{eq:Path Integral Expression of Partition Function}
\end{eqnarray}
with
\begin{eqnarray}
S[\bar{\psi},\psi] &\equiv& \sum_{m=-\infty}^{\infty} \sum_{j,j'} \sum_{\bm{k}} \bar{\psi}_{\bm{k},j,m} \Big(A_{jj'}(\bm{k}) - (\mu+i\omega_m) \delta_{jj'}\Big) \psi_{\bm{k},j',m} ,
\end{eqnarray}
and 
\begin{eqnarray}
\int_{-\infty}^{\infty} \mathcal{D} (\bar{\psi},\psi) \equiv \int_{-\infty}^{\infty}  \Big(\prod_{\bm{k}} \prod_{j=1}^{L} \prod_{m=-\infty}^{\infty} \frac{d\bar{\psi}_{\bm{k},j,m} d\psi_{\bm{k},j,m}}{\beta \pi} \Big) ,
\end{eqnarray}
where $\omega_m$ is the Matsubara frequency in Boson systems
\begin{eqnarray}
\omega_m \equiv \frac{2 m \pi}{\beta} .
\end{eqnarray}
Because the integral in (\ref{eq:Path Integral Expression of Partition Function}) is Gaussian, the convergence condition of this integral is that all eigenvalues of the $L\times L$ matrix $A(\bm{k}) - (\mu+i\omega_m) E_{L}$ have a positive real part. This condition is the same as (\ref{eq:range_mu}).

When the condition (\ref{eq:range_mu}) is satisfied, we can calculate the integral in (\ref{eq:Path Integral Expression of Partition Function}) as
\begin{eqnarray}
e^{- \beta J(\beta,\mu)} = \prod_{m=-\infty}^{\infty} \prod_{\bm{k}} \frac{\beta^{-L}}{\det \Big(A(\bm{k}) - (\mu+i\omega_m) E_{L} \Big)} ,
\label{eq:J_beforeSumMat}
\end{eqnarray}
which leads to
\begin{eqnarray}
\beta J(\beta,\mu) = \sum_{m=-\infty}^{\infty} \sum_{\bm{k}} \log \Big[\beta^L \det \Big(A(\bm{k}) - (\mu+i\omega_m) E_{L} \Big)\Big] .
\end{eqnarray}
Computing the sum over the Matsubara frequencies, we obtain (\ref{eq:J}).

\subsection{Derivation of (\ref{eq:path integral calculation u1n2})}
\label{sec:derivation of (u1n2)}
To derive (\ref{eq:path integral calculation u1n2}), we start with the path integral expression of $<\sum_{\bm{k}} \hat{a}^{\dagger}_{\bm{k},j} \hat{a}_{\bm{k},j'} >_{\beta,\mu}$:
\begin{eqnarray}
<\sum_{\bm{k}} \hat{a}^{\dagger}_{\bm{k},j} \hat{a}_{\bm{k},j'} >_{\beta,\mu} &=& e^{\beta J(\beta,\mu)} \sum_{n} <\bm{n}|\sum_{\bm{k}} \hat{a}^{\dagger}_{\bm{k},j} \hat{a}_{\bm{k},j'} e^{-\beta(\hat{H}-\mu \hat{N})}|\bm{n}> \nonumber \\[3pt]
&=&e^{\beta J(\beta,\mu)}  \frac{1}{\beta}\sum_{\bm{k}} \sum_{m,m'}^{} \int_{-\infty}^{\infty} \mathcal{D} (\bar{\psi},\psi) \bar{\psi}_{\bm{k},j}^{m} \psi_{\bm{k},j'}^{m'}  e^{-S[\bar{\psi},\psi]} . \qquad \nonumber\\
\label{eq:<aa>}
\end{eqnarray}
When the condition (\ref{eq:range_mu}) is satisfied, we can calculate the integral in (\ref{eq:<aa>}) as
\begin{eqnarray} 
\int_{-\infty}^{\infty} \mathcal{D} (\bar{\psi},\psi) \bar{\psi}_{\bm{k},j}^{m} \psi_{\bm{k},j'}^{m'}  e^{-S[\bar{\psi},\psi]} = e^{-\beta J(\beta,\mu)} \Big(A(\bm{k}) - (\mu+i\omega_m) E_{L} \Big)_{jj'}^{-1} \delta_{m,m'} . \qquad\nonumber\\
\end{eqnarray}
Substituting this result into (\ref{eq:<aa>}), we obtain
\begin{eqnarray}
<\sum_{\bm{k}} \hat{a}^{\dagger}_{\bm{k},j} \hat{a}_{\bm{k},j'} >_{\beta,\mu} =   \frac{1}{\beta}\sum_{\bm{k}} \sum_{m=-\infty}^{\infty} \Big(A(\bm{k}) - (\mu+i\omega_m) E_{L} \Big)_{jj'}^{-1} .
\label{eq:res_<aa>}
\end{eqnarray}

Here, we focus on the component $(j,j')=(1,1)$ and we rewrite the inverse of the matrix $A(\bm{k}) - (\mu+i\omega_m) E_{L}$ as
\begin{eqnarray}
\Big(A(\bm{k}) - z E_{L} \Big)_{11}^{-1}&=& \frac{\det \Big(B_{L-1}(\bm{k}) - z E_{L-1} \Big)}{\det \Big(A(\bm{k}) -z E_{L} \Big)} \nonumber \\
&=& \frac{\det \Big(B_{L-1}(\bm{k}) - z E_{L-1} \Big)}{\prod_{n=1}^{N_b} \Big(-z+\epsilon_n(\bm{k}) \Big)} .
\label{eq:(A-zE)-1}
\end{eqnarray}
Using (\ref{eq:(A-zE)-1}), we can formally compute the sum over the Matsubara frequencies of (\ref{eq:res_<aa>}) as follows
\begin{eqnarray}
& & \frac{1}{\beta} \sum_{m=-\infty}^{\infty} \Big(A(\bm{k}) - (\mu+i\omega_m) E_{L} \Big)_{11}^{-1} \nonumber \\[3pt]
&=& \frac{1}{2\pi i} \oint_{\gamma} dz \ \Big(A(\bm{k}) - (\mu+z) E_{L} \Big)_{11}^{-1} \frac{1}{e^{\beta z}-1}\nonumber \\[3pt]
&=& \frac{1}{2\pi i} \oint_{\gamma} dz  \frac{\det \Big(B_{L-1}(\bm{k}) - (\mu + z) E_{L-1} \Big)}{\prod_{n=1}^{L} \Big(-z - \mu +\epsilon_n(\bm{k}) \Big)} \frac{1}{e^{\beta z}-1} \nonumber \\[3pt]
&=& - \sum_{n=1}^{L} \lim_{z \to - \mu + \epsilon_n(\bm{k})} \biggl\{(-z-\mu+\epsilon_n(\bm{k}))\frac{\det \Big(B_{L-1}(\bm{k}) - (\mu + z) E_{L-1} \Big)}{\prod_{n=1}^{L} \Big(-z - \mu +\epsilon_n(\bm{k}) \Big)} \frac{1}{e^{\beta z}-1} \biggr\}  \nonumber \\
&=& \sum_{n=1}^{L} \frac{f_n(\bm{k})}{e^{\beta(\epsilon_n(\bm{k})-\mu)}-1} ,
\end{eqnarray}
where we choose the integration contour $\gamma$ so as to enclose the poles $\{i\omega_m\}_{m}$ in the clockwise direction and we introduce $f_n(\bm{k})$ as
\begin{eqnarray}
f_n(\bm{k}) \equiv \lim_{z \to \epsilon_n(\bm{k})} \biggl\{(-z + \epsilon_n(\bm{k})) \frac{\det \Big(B_{L-1}(\bm{k}) - z E_{L-1} \Big)}{\det \Big(A(\bm{k}) -z E_{L} \Big)} \biggr\} .
\end{eqnarray}
Recalling (\ref{eq:Nj after calculation}), we obtain
\begin{eqnarray}
<\sum_{\bm{k}} \hat{a}^{\dagger}_{\bm{k},1} \hat{a}_{\bm{k},1} > &=& \sum_{\bm{k}} \sum_{n=1}^L |u_n^1(\bm{k})|^2  \frac{1}{e^{\beta(\epsilon_n(\bm{k})-\mu)}-1}.
\end{eqnarray}
Therefore, we obtain
\begin{eqnarray}
|u^1_n(\bm{k})|^2 = f_n(\bm{k}) .
\end{eqnarray}

\subsection{Derivation of the effective surface system}
\label{sec:derivation of the effective surface system}
We obtain the effective action associated with the surface system by integrating (\ref{eq:Path Integral Expression of Partition Function}) over the degrees of the freedom $(\psi_{\bm{k},n,m},\bar{\psi}_{\bm{k},n,m})_{\bm{k},n\geq 2,m}$ except for the degrees of the freedom $(\psi_{\bm{k},1,m},\bar{\psi}_{\bm{k},1,m})_{\bm{k},m}$. 

In order to compute this procedure, we pick up one component $(\bm{k},m)$ from $S[\bar{\psi},\psi]$ and define as
\begin{eqnarray}
S(\bm{\bar{\psi}}_{\bm{k},m},\bm{\psi}_{\bm{k},m}) \equiv \sum_{j,j'} \bar{\psi}_{\bm{k},j,m} \Big(A_{jj'}(\bm{k}) - (\mu+i\omega_m) \delta_{jj'}\Big) \psi_{\bm{k},j',m} .
\end{eqnarray}
%it is better to be split $S(\bm{\bar{\psi}}_{\bm{k},m},\bm{\psi}_{\bm{k},m})$ into three term as 
%\begin{eqnarray}
%S[\bar{\psi},\psi] &=& \sum_{j=2}^{L} \sum_{j'=2}^{L} \bar{\psi}_{\bm{k},j,m} \Big(A_{jj'}(\bm{k}) - (\mu+i\omega_m) \delta_{jj'}\Big) \psi_{\bm{k},j',m} \nonumber \\[3pt]
%&+& \Big( \bar{\psi}_{\bm{k},1,m} A_{12}(\bm{k}) \psi_{\bm{k},2,m} + \bar{\psi}_{\bm{k},2,m} A_{21}(\bm{k}) \psi_{\bm{k},1,m} \Big) \nonumber \\[3pt]
%&+& \bar{\psi}_{\bm{k},1,m} A_{11}(\bm{k}) \psi_{\bm{k},1,m}
%\end{eqnarray}
By straightforward calculation, we obtain
\begin{eqnarray}
& &\int_{-\infty}^{\infty}  \Big( \prod_{i=2}^{L} \frac{d\bar{\psi}_{\bm{k},i,m} d\psi_{\bm{k},i,m}}{\beta \pi} \Big) e^{-S(\bm{\bar{\psi}}_{\bm{k},m},\bm{\psi}_{\bm{k},m})} \nonumber \\[3pt]
&=& \frac{\beta^{-(L-1)}}{\det \Big(B_{L-1}(\bm{k})-(\mu+i\omega_m)E_{L-1}\Big)}e^{-S_{1}(\bar{\psi}_{\bm{k},1,m},\psi_{\bm{k},1,m};\bm{k},m)} ,
\label{eq:Integral Path Integral Part 1}
\end{eqnarray}
where
\begin{eqnarray}
S_{1}(\bar{\psi}_{\bm{k},1,m},\psi_{\bm{k},1,m};\bm{k},m) \equiv \bar{\psi}_{\bm{k},1,m} \mathcal{G}(\bm{k},\omega_m;\mu) \psi_{\bm{k},1,m}
\end{eqnarray}
with
\begin{eqnarray}
\mathcal{G}(\bm{k},\omega_m;\mu) \equiv A_{11}(\bm{k})-(\mu+i\omega_m) - J^2 \Big(B_{L-1}(\bm{k})- (\mu+i\omega_m) E_{L-1} \Big)^{-1}_{11} . \qquad
\end{eqnarray}
Using (\ref{eq:Integral Path Integral Part 1}), we integrate (\ref{eq:Path Integral Expression of Partition Function}) over the degrees of freedom $(\psi_{\bm{k},n,m},\bar{\psi}_{\bm{k},n,m})_{\bm{k},n\geq 2,m}$ as
\begin{eqnarray}
e^{-\beta J(\beta,\mu)} &=& e^{-\beta J_{L-1}(\beta,\mu;\Delta=0)} \int_{-\infty}^{\infty} \mathcal{D} (\bar{\psi}_1,\psi_1) e^{-S_1[\bar{\psi}_1,\psi_1]}
\end{eqnarray}
with
\begin{eqnarray}
S_1[\bar{\psi}_1,\psi_1] \equiv \sum_{\bm{k},m} S_{1}(\bar{\psi}_{\bm{k},1,m},\psi_{\bm{k},1,m};\bm{k},m) ,
\end{eqnarray}
\begin{eqnarray}
\int_{-\infty}^{\infty} \mathcal{D} (\bar{\psi}_1,\psi_1) \equiv \int_{-\infty}^{\infty}  \Big(\prod_{\bm{k}} \prod_{m=-\infty}^{\infty} \frac{d\bar{\psi}_{\bm{k},1,m} d\psi_{\bm{k},1,m}}{\beta \pi} \Big),
\end{eqnarray}
and
\begin{eqnarray}
e^{-\beta J_{L-1}(\beta,\mu;\Delta=0)} = \prod_{m=-\infty}^{\infty} \prod_{\bm{k}} \frac{\beta^{-(L-1)}}{\det \Big(B_{L-1}(\bm{k})-(\mu+i\omega_m)E_{L-1}\Big)} ,
\end{eqnarray}
where we have used (\ref{eq:J_beforeSumMat}). Note that $J_{L-1}(\beta,\mu;\Delta=0)$ corresponds to the total free energy of the system consisting of $L-1$ layers with $\Delta=0$.

As a special case, we focus on $\mathcal{G}(\bm{k},0;\mu)$. Recalling that the eigenvalues and eigenvectors of the $L \times L$ matrix $B(\bm{k})$ are given by (\ref{eq:Eigenvalue B}) and (\ref{eq:Eigenvector B}) respectively, we can calculate
\begin{eqnarray}
& & \Big(B_{L-1}(\bm{k})-(\mu+i\omega_m)E_{L-1} \Big)^{-1}_{11} \nonumber \\[3pt]
&=& \frac{2}{L-2} \sum_{l=1}^{L-1} \frac{\sin^2 (\frac{l\pi}{L})}{\omega(\bm{k})-(\mu+i\omega_m) - 2t \cos(\frac{l\pi}{L})} \nonumber \\[3pt]
&\simeq& 2 \int_{0}^{2\pi} \frac{d\theta}{2\pi} \frac{\sin^2 \theta}{\omega(\bm{k})-(\mu+i\omega_m) - 2t \cos \theta}
\label{eq:(B-zE)-1_11}
\end{eqnarray}
for any $m$, where we have taken the large system size limit for the last equation. For the case $m=0$, this integral is calculated as
\begin{eqnarray}
\int_{0}^{2\pi} \frac{d\theta}{2\pi} \frac{\sin^2 \theta}{\omega(\bm{k})-\mu - 2t \cos \theta} =\frac{(\omega(\bm{k})-\mu)-\sqrt{(\omega(\bm{k})-\mu)^2-(2t)^2}}{(2t)^2} ,
\end{eqnarray}
where $\mu<-6t$. Using this result, we obtain
%\begin{eqnarray}
%& & \lim_{L\to\infty}\Big((A_{11}(\bm{k})-\mu) -  J^2(B_{L-1}(\bm{k})- \mu E_{L-1})^{-1}_{11} \Big) \nonumber \\[3pt]
%&=&  (1+\Delta)\omega(\bm{k}) - \mu - \frac{(\omega(\bm{k})-\mu)-\sqrt{(\omega(\bm{k})-\mu)^2-(2J)^2}}{2} \nonumber \\
%&=& \frac{(1+2\Delta)\omega(\bm{k})-\mu + \sqrt{(\omega(\bm{k})-\mu)^2-(2J)^2}}{2}
%\end{eqnarray}
\begin{eqnarray}
\lim_{L \to \infty} \mathcal{G}(\bm{k},0;\mu) = \frac{(1+2\Delta)\omega(\bm{k})-\mu + \sqrt{(\omega(\bm{k})-\mu)^2-(2t)^2}}{2} ,
\label{eq:G(0)}
\end{eqnarray}
where  $\mu<-6t$.

\section{Derivation of (\ref{eq:eq_mu_gen})}
\label{sec:derivation of (66)}
We derive (\ref{eq:eq_mu_gen}) from (\ref{eq:eq_mu}) in the thermodynamic limit. Especially we focus on the singularity when $\mu$ approaches the value $\epsilon_1(\bm{0})$. In order to see it, we divide the right hand side of (\ref{eq:eq_mu}) into four terms as
\begin{eqnarray}
\rho &=&  \frac{1}{L^3} \frac{1}{e^{\beta (\epsilon_1(\bm{0})-\mu)}-1}  +  \frac{1}{L^3} \sum_{\substack{\bm{k} (\neq \bm{0}) \\ (\omega(\bm{k})<0)}} \frac{1}{e^{\beta (\epsilon_1(\bm{k})-\mu)}-1} \nonumber \\[3pt]
&+&  \frac{1}{L^3} \sum_{n=2}^{L} \sum_{\substack{\bm{k} \\ (\omega(\bm{k})<0)}} \frac{1}{e^{\beta (\epsilon_n(\bm{k})-\mu)}-1} +   \frac{1}{L^3} \sum_{n=1}^{L} \sum_{\substack{\bm{k} \\ (\omega(\bm{k})\geq0)}} \frac{1}{e^{\beta (\epsilon_n(\bm{k})-\mu)}-1} .\quad
\label{eq:sep_eqmu}
\end{eqnarray}
It should be noted that $\epsilon_1(\bm{k})$ in $\Delta>\frac{1}{4}$ finitely deviates from that of $\Delta=0$ in some $\bm{k}$ regime.

\subsection{Preliminary}
As a preliminary, we estimate $\epsilon_1(\bm{k}') - \epsilon_1(\bm{0})$ with $\bm{k}'=(2\pi/L,0)$ and $\epsilon_2(\bm{0}) - \epsilon_1(\bm{0})$ in the large system size limit.
First we consider the case of $\Delta < 1/4$. Substituting (\ref{eq:Definition Omega(k)}) into (\ref{eq:eq_THETA}) and using addition formulas, we obtain
\begin{eqnarray}
\tan (L\theta) &=& -\frac{\sin \theta}{\cos \theta -2 \Delta \sum_{d=1}^2 \cos k_d} \nonumber \\[3pt]
&\simeq& - \frac{\sin \theta}{\cos \theta - (4 \Delta - \Delta |\bm{k}|^2)},
\label{eq:eq_THETA after deform}
\end{eqnarray}
where $\bm{k}$ is sufficiently small in the second line.
\begin{figure}
\centering
\includegraphics[width=8cm]{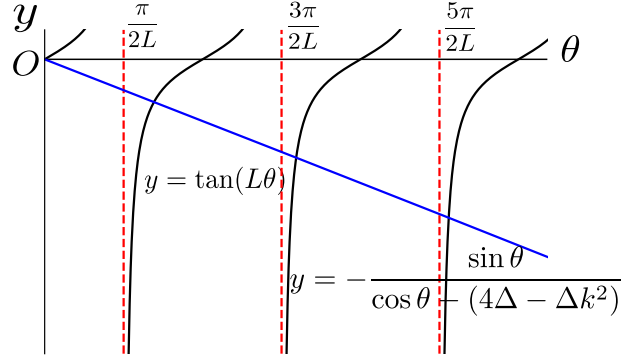}
\caption{Schematic graph of $y=\tan(L\theta)$ and $y=- \sin \theta/(\cos \theta - (4 \Delta - \Delta |\bm{k}|^2))$. Intersections of these graphs correspond to solutions of (\ref{eq:eq_THETA after deform}).}
\label{fig:tan(Ltheta)}
\end{figure}
To estimate the solutions of (\ref{eq:eq_THETA after deform}), we use Figure.~\ref{fig:tan(Ltheta)}. In Figure.~\ref{fig:tan(Ltheta)}, the points of intersection of two graphs correspond to the solutions of (\ref{eq:eq_THETA after deform}). Therefore we find
\begin{eqnarray}
\frac{\pi}{2L} < \theta_1(\bm{0}) < \frac{\pi}{L} ,
\label{eq:theta condition1}
\end{eqnarray}
and
\begin{eqnarray}
\frac{3\pi}{2L} < \theta_2(\bm{0}) < \frac{2\pi}{L}.
\label{eq:theta condition2}
\end{eqnarray} 
Futhermore, because
\begin{eqnarray}
- \frac{\sin \theta}{\cos \theta - 4 \Delta} < - \frac{\sin \theta}{\cos \theta - (4 \Delta - \Delta |\bm{k}'|^2)}
\end{eqnarray}
near $\theta = 0$, we find
\begin{eqnarray}
\frac{\pi}{2L} < \theta_1(\bm{0}) < \theta_1(\bm{k}') < \frac{\pi}{L}.
\label{eq:theta condition3}
\end{eqnarray}
From (\ref{eq:eq_EE}), (\ref{eq:theta condition1}), (\ref{eq:theta condition2}) and (\ref{eq:theta condition3}), we obtain
\begin{eqnarray}
\epsilon_2(\bm{0}) - \epsilon_1(\bm{0}) = O(\frac{1}{L^2}),
\label{eq:eps2-eps1 inside}
\end{eqnarray}
and
\begin{eqnarray}
\epsilon_1(\bm{k}') - \epsilon_1(\bm{0}) = O(\frac{1}{L^2}).
\label{eq:eps1'-eps1 inside}
\end{eqnarray}
Next, we consider the case of $\Delta \geq 1/4$. Because (\ref{eq:eps1 in the large system size limit outside}) implies 
\begin{eqnarray}
\epsilon_1(\bm{k}')-\epsilon_1(\bm{0}) \simeq |\bm{k}'|^2,
\end{eqnarray}
it is reasonable to conjecture
\begin{eqnarray}
\epsilon_1(\bm{k}') - \epsilon_1(\bm{0}) = O(\frac{1}{L^2}).
\label{eq:eps1'-eps1 outside}
\end{eqnarray}
From (\ref{eq:rela_EE1}) and (\ref{eq:stronger condition for eps1 in low Delta}), we obtain
\begin{eqnarray}
(\omega(\bm{0})-2t) - \epsilon_1(\bm{0}) < \epsilon_2(\bm{0}) - \epsilon_1(\bm{0}) ,
\label{eq:eps2-eps1 outside}
\end{eqnarray}
and from (\ref{eq:Definition Omega(k)}) and (\ref{eq:eps1 in the large system size limit outside}), we obtain
\begin{eqnarray}
\lim_{L \to \infty} \{(\omega(\bm{0})-2t) - \epsilon_1(\bm{0}) \} = t \frac{1+16\Delta^2}{4\Delta} - 2t.
\label{eq:eps2-eps1 outside ap}
\end{eqnarray}
From (\ref{eq:eps2-eps1 outside}) and (\ref{eq:eps2-eps1 outside ap}), we find that the difference between $\epsilon_1(\bm{0})$ and $\epsilon_2(\bm{0})$ is the finite. 
To summarize these results, the energy gap between the grand state and the first excited state is always $O(L^{-2})$ in the large system size limit.

\subsection{Third term of (\ref{eq:sep_eqmu}) in the thermodynamic limit}
Based on the preliminary results, we consider the third term of (\ref{eq:sep_eqmu}). From (\ref{eq:rela_EE1}) and (\ref{eq:rela_EE}), we find that $\epsilon_n(\bm{k})$ does not finitely deviate from that for $\Delta=0$ in any $\Delta$. As the result, the third term of (\ref{eq:sep_eqmu}) is the same as that of $\Delta=0$ in the thermodynamic limit. We demonstrate this.

Using (\ref{eq:rela_EE1}) and (\ref{eq:rela_EE}), we have
\begin{eqnarray}
\frac{1}{e^{\beta (\epsilon^0_n(\bm{k})-\mu)}-1} < \frac{1}{e^{\beta (\epsilon_n(\bm{k})-\mu)}-1} <\frac{1}{e^{\beta (\epsilon^0_{n-1}(\bm{k})-\mu)}-1}
\label{eq:inequality omega(k)<0}
\end{eqnarray}
for $n=2,3,\cdots,L$, where $\bm{k}$ satisfies $\omega(\bm{k})<0$. (\ref{eq:inequality omega(k)<0}) immediately leads to
\begin{eqnarray}
& &\frac{1}{L^3} \sum_{n=2}^{L} \sum_{\substack{\bm{k} \\ (\omega(\bm{k})<0)}} \frac{1}{e^{\beta (\epsilon^0_n(\bm{k})-\mu)}-1} < \frac{1}{L^3} \sum_{n=2}^{L} \sum_{\substack{\bm{k} \\ (\omega(\bm{k})<0)}}  \frac{1}{e^{\beta (\epsilon_n(\bm{k})-\mu)}-1} \nonumber \\[3pt]
&<& \frac{1}{L^3} \sum_{n=2}^{L} \sum_{\substack{\bm{k} \\ (\omega(\bm{k})<0) \\({\rm except\  for \ } \epsilon^0_1(\bm{0}))}}  \frac{1}{e^{\beta (\epsilon^0_{n-1}(\bm{k})-\mu)}-1} +  \frac{1}{L^3} \frac{1}{e^{\beta (\epsilon_{2}(\bm{0})-\mu)}-1} .
\label{eq:inequality about third term}
\end{eqnarray}
Using (\ref{eq:range_mu}) we evaluate the last term in this inequality as
\begin{eqnarray}
\frac{1}{L^3} \frac{1}{e^{\beta (\epsilon_2(\bm{0}) - \mu)}-1} < \frac{1}{L^3} \frac{1}{e^{\beta (\epsilon_2(\bm{0}) - \epsilon_1(\bm{0}))}-1} \sim O(\frac{1}{L}),
\label{eq:the last term in inequality about third term}
\end{eqnarray}
where we have used (\ref{eq:eps2-eps1 inside}) and (\ref{eq:eps2-eps1 outside}). Using (\ref{eq:inequality about third term}) and (\ref{eq:the last term in inequality about third term}), we confirm
\begin{eqnarray}
\lim_{L \to \infty} \frac{1}{L^3} \sum_{n=2}^{L} \sum_{\substack{\bm{k} \\ (\omega(\bm{k})<0)}} \frac{1}{e^{\beta (\epsilon_n(\bm{k})-\mu)}-1} =  \int_{\omega(\bm{k})<0} \frac{d^3 \bm{\rho}}{(2\pi)^2 \pi} \frac{1}{e^{\beta (-2t \sum_{d=1}^3 \cos \rho_d - \mu)}-1},
\label{eq: third sum to integral one part}
\end{eqnarray}
%where we have used
%\begin{eqnarray}
%& & \lim_{L \to \infty} \frac{1}{L^3} \sum_{\substack{\bm{k} \\ (\omega(\bm{k})<0) \\ ({\rm except\  for \ } \epsilon^0_1(\bm{0}))}} \frac{1}{e^{\beta (\epsilon^0_{1}(\bm{k})-\mu)}-1} \nonumber \\[3pt]
%& & \qquad \qquad = \lim_{L \to \infty} \frac{1}{L} \int_{\omega(\bm{k})<0} \frac{d^2 \bm{k}}{(2\pi)^2} \frac{1}{e^{\beta (-J \sum_{d=1}^2 \cos k_d - \mu)}-1} = 0
%\end{eqnarray}
where $\bm{\rho} = (k_1,k_2,\rho_3)$.

\subsection{Fourth term of (\ref{eq:sep_eqmu}) in the thermodynamic limit}
Next, we consider the second term of (\ref{eq:sep_eqmu}). Using (\ref{eq:rela_EE1}), (\ref{eq:rela_EE}) and (\ref{eq:relation k and k'}) we have 
\begin{eqnarray}
0 < \frac{1}{e^{\beta (\epsilon_{L}(\bm{k}')-\mu)}-1} <\frac{1}{e^{\beta (\epsilon^0_{L}(\bm{k}')-\mu)}-1},
\label{eq:inequality omega(k)>0 1}
\end{eqnarray}
and 
\begin{eqnarray}
\frac{1}{e^{\beta (\epsilon^0_{n+1}(\bm{k'})-\mu)}-1} < \frac{1}{e^{\beta (\epsilon_{n}(\bm{k}')-\mu)}-1} <\frac{1}{e^{\beta (\epsilon^0_{n}(\bm{k}')-\mu)}-1}
\label{eq:inequality omega(k)>0 2}
\end{eqnarray}
for $n=1,2,\cdots,L-1$, where $\bm{k}'$ satisfies $\omega(\bm{k}')>0$. (\ref{eq:inequality omega(k)>0 1}) and (\ref{eq:inequality omega(k)>0 2}) lead to
\begin{eqnarray}
& &\frac{1}{L^3} \sum_{n=1}^{L-1} \sum_{\substack{\bm{k}' \\ (\omega(\bm{k}')>0)}} \frac{1}{e^{\beta (\epsilon^0_{n+1}(\bm{k}')-\mu)}-1} < \frac{1}{L^3} \sum_{n=1}^{L} \sum_{\substack{\bm{k}' \\ (\omega(\bm{k}')>0)}}  \frac{1}{e^{\beta (\epsilon_{n}(\bm{k}')-\mu)}-1} \nonumber \\[3pt]
&<& \frac{1}{L^3} \sum_{n=1}^{L} \sum_{\substack{\bm{k}' \\ (\omega(\bm{k}')>0)}}  \frac{1}{e^{\beta (\epsilon^0_{n}(\bm{k}')-\mu)}-1} .
\label{eq:inequality about third term 2}
\end{eqnarray}
From (\ref{eq:eps_n=eps_n0}) and (\ref{eq:inequality about third term 2}) we obtain
\begin{eqnarray}
\lim_{L \to \infty} \frac{1}{L^3} \sum_{n=1}^{L} \sum_{\substack{\bm{k}' \\ (\omega(\bm{k}')\geq0)}}  \frac{1}{e^{\beta (\epsilon_{n}(\bm{k}')-\mu)}-1} = \int_{\omega(\bm{k}')\geq0} \frac{d^3 \bm{\rho}}{(2\pi)^2 \pi} \frac{1}{e^{\beta (-2t \sum_{d=1}^3 \cos \rho_d - \mu)}-1}.
\label{eq: third sum to integral another part}
\end{eqnarray}
Combining (\ref{eq: third sum to integral one part}) and (\ref{eq: third sum to integral another part}) we obtain
\begin{eqnarray}
& & \lim_{L \to \infty} \Big[\frac{1}{L^3} \sum_{n=1}^{L} \sum_{\substack{\bm{k} \\ (\omega(\bm{k})\geq0)}} \frac{1}{e^{\beta (\epsilon_{n}(\bm{k})-\mu)}-1} + \frac{1}{L^3} \sum_{n=2}^{L} \sum_{\substack{\bm{k} \\ (\omega(\bm{k})<0)}} \frac{1}{e^{\beta (\epsilon_n(\bm{k})-\mu)}-1} \Big] \nonumber \\[3pt]
&=& \int \frac{d^3 \bm{\rho}}{(2\pi)^2 \pi} \frac{1}{e^{\beta (-2t \sum_{d=1}^3 \cos \rho_d - \mu)}-1}.
\label{eq: third sum to integral complete part}
\end{eqnarray}

\subsection{Second term of (\ref{eq:sep_eqmu}) in the thermodynamic limit}
Finally, we consider the second term of (\ref{eq:sep_eqmu}). Using (\ref{eq:range_mu}), we obtain
\begin{eqnarray}
\frac{1}{L^3} \sum_{\substack{\bm{k} (\neq \bm{0})\\ (\omega(\bm{k})<0)}} \frac{1}{e^{\beta (\epsilon_1(\bm{k})-\mu)}-1} < \frac{1}{L^3} \sum_{\substack{\bm{k} (\neq \bm{0}) \\ (\omega(\bm{k})<0)}} \frac{1}{e^{\beta (\epsilon_1(\bm{k})-\epsilon_1(\bm{0}))}-1}.
\label{eq:inequality about the second term}
\end{eqnarray}
As $L \to \infty$, the summation in the right hand side of (\ref{eq:inequality about the second term}) can be replaced by the integral
\begin{eqnarray}
& & \lim_{L\to \infty} \frac{1}{L^3} \sum_{\substack{\bm{k} (\neq \bm{0}) \\ (\omega(\bm{k})<0)}} \frac{1}{e^{\beta (\epsilon_1(\bm{k})-\epsilon_1(\bm{0}))}-1} \nonumber \\[3pt]
&=& \lim_{L \to \infty} \frac{1}{L} \int_{\frac{2\pi}{L}}^{} \frac{d^2 \bm{k}}{(2\pi)^2} \frac{1}{e^{\beta (\epsilon_1(\bm{k})-\epsilon_1(\bm{0}))}-1},
\label{eq:eqmu2nd}
\end{eqnarray}
where we have used (\ref{eq:eps1'-eps1 inside}). We devide this integral by introducing a small finite wavelength $\Lambda$ as
\begin{eqnarray}
\int_{2\pi/L} \frac{d^2 \bm{k}}{(2\pi)^2} \frac{1}{e^{\beta(\epsilon_1(\bm{k}) - \epsilon_1(\bm{0}))}-1} &=& \int_{2\pi/L}^{\Lambda} \frac{d^2 \bm{k}}{(2\pi)^2} \frac{1}{e^{\beta(\epsilon_1(\bm{k}) - \epsilon_1(\bm{0}))}-1} \nonumber \\[3pt]
&+& \int_{\Lambda}^{} \frac{d^2 \bm{k}}{(2\pi)^2} \frac{1}{e^{\beta(\epsilon_1(\bm{k}) - \epsilon_1(\bm{0}))}-1}.
\label{eq:sep_eqmu2nd}
\end{eqnarray}
The second term converges, while the first term diverges because we estimate 
\begin{eqnarray}
\int_{2\pi/L}^{\Lambda} \frac{d^2 \bm{k}}{(2\pi)^2} \frac{1}{e^{\beta(\epsilon_1(\bm{k}) - \epsilon_1(\bm{0}))}-1} &\sim&  \int_{2\pi/L}^{\Lambda} \frac{d^2 \bm{k}}{(2\pi)^2} \frac{1}{\beta(\epsilon_1(\bm{k})-\epsilon_1(\bm{0}))} \nonumber \\[3pt]
&\sim&  \int_{2\pi/L}^{\Lambda} \frac{d^2 \bm{k}}{(2\pi)^2} \frac{1}{\beta k^2} \nonumber \\[3pt]
&\sim& O(\log L).
\end{eqnarray}
Because this divergence is $O(\log L)$,  (\ref{eq:eqmu2nd}) becomes
\begin{eqnarray}
\lim_{L\to \infty} \frac{1}{L^3} \sum_{\bm{k}(\neq \bm{0})} \frac{1}{e^{\beta (\epsilon_1(\bm{k})-\mu)}-1} = 0.
\label{eq:second sum to integral}
\end{eqnarray}
As the result, the second term of (\ref{eq:sep_eqmu}) can be neglected. By combining (\ref{eq: third sum to integral complete part}) and (\ref{eq:second sum to integral}), we obtain (\ref{eq:eq_mu_gen}).

%%%%%%%%%%%%%%%%%%%%%%%
% References          %
%%%%%%%%%%%%%%%%%%%%%%%

\end{document}